\documentclass[ras]{agutex}
\usepackage{amsmath}
\usepackage[dvips]{graphicx}
\usepackage[dvips]{color}
\setkeys{Gin}{draft=false}
\usepackage{amsmath}
\usepackage{amssymb}

 \usepackage{lineno}
 \nolinenumbers       

\authorrunninghead{SUTINJO ET AL.}

\titlerunninghead{UNDERSTANDING STOKES LEAKAGE IN MWA}

\authoraddr{Corresponding author: A. Sutinjo,
Curtin Institute of Radio Astronomy, International Centre for Radio Astronomy Research (ICRAR)/Curtin University, 1 Turner Avenue, Bentley WA 6102.
(adrian.sutinjo@icrar.org)}

\begin{document}

%
%

\title{Understanding Instrumental Stokes Leakage in Murchison Widefield Array Polarimetry}
%
%

%
%



 \authors{A. Sutinjo\altaffilmark{1},
 J. O'Sullivan\altaffilmark{2,3}, E. Lenc\altaffilmark{4,5},
R. B. Wayth\altaffilmark{1}, S. Padhi\altaffilmark{1}, P. Hall\altaffilmark{1,2} and S. J. Tingay\altaffilmark{1}}

\altaffiltext{1}{International Centre for Radio Astronomy Research (ICRAR)/Curtin University, Bentley, Western Australia}

\altaffiltext{2}{Department of Electrical and Computer Engineering, Curtin University, Bentley, Western Australia}

\altaffiltext{3}{Formerly CSIRO Centre Astronomy and Space Science (CASS), Australia}

\altaffiltext{4}{Sydney Institute for Astronomy, School of Physics, The University of Sydney, NSW, Australia}

\altaffiltext{5}{ARC Centre of Excellence for All-sky Astrophysics (CAASTRO), Sydney, Australia }



%
%

\begin{abstract}
This paper offers an electromagnetic, more specifically array theory, perspective on understanding strong instrumental polarization effects for planar low-frequency ``aperture arrays'' with the Murchison Widefield Array (MWA) as an example. A long-standing issue that has been seen here is significant instrumental Stokes leakage after calibration, particularly in Stokes Q at high frequencies. A simple model that accounts for inter-element mutual coupling is presented which explains the prominence of Q leakage seen when the array is scanned away from zenith in the principal planes. On these planes, the model predicts current imbalance in the X (E-W) and Y (N-S) dipoles and hence the Q leakage. Although helpful in concept, we find that this  model is inadequate to explain the full details of the observation data. This finding motivates further experimentation with more rigorous models that account for both mutual coupling and embedded element patterns. Two more rigorous models are discussed: the ``full'' and ``average'' embedded element patterns. The viability of the ``full'' model is demonstrated by simulating current MWA practice of using a Hertzian dipole model as a Jones matrix estimate. We find that these results replicate the observed Q leakage to approximately 2~to~5\%. Finally, we offer more direct indication for the level of improvement expected from upgrading the Jones matrix estimate with more rigorous models. Using the ``average'' embedded pattern as an estimate for the ``full'' model, we find that Q leakage of a few percent is achievable.

\end{abstract}

%
%

%

\begin{article}

%
%

\section{Introduction}
\label{sec:Intro}
\begin{figure}[t]
	\begin{center}
	{\includegraphics[width=3in]{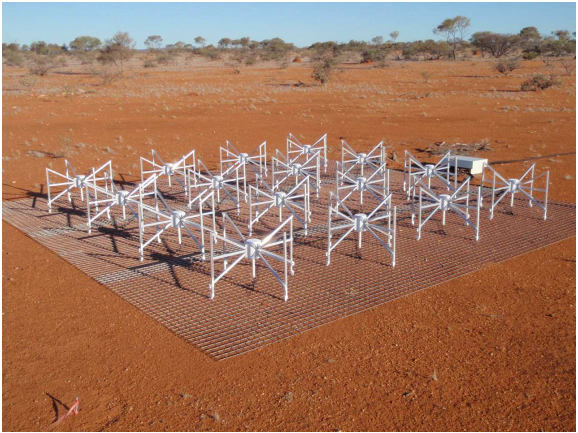}}
	\end{center}
\caption{A photo of an MWA tile. Each tile is comprised of  4$\times$4 bow-tie antennas regularly spaced at 1.1~m on a 5$\times$5~m wire grid. The bow-tie arms are oriented E-W and N-S. Beam-scanning is accomplished by switching analog delay-lines (contained in the white box next to the metallic grid in the photo) commensurate with pointing directions. }
\label{fig:Tile}
\end{figure}

	The Murchison Widefield Array (MWA) is a precursor to the Square Kilometre Array (SKA) low-frequency aperture arrays (LFAA) situated at the Murchison Radio-astronomy Observatory (MRO) in the mid-west of Western Australia~\cite{Lonsdale_2009, 2013PASA...30....7T, RDS:RDS6017}. The MWA operates in the 80-300 MHz band and consists of 128 aperture array ``tiles'' \textcolor{black}{(Fig.~\ref{fig:Tile})} spread over an area of approximately 3~km. It is closely related in frequency band to SKA LFAA pathfinders such LOFAR~\cite{refId0, RDS:RDS5907} in Europe and LWA in New Mexico, USA.~\cite{ 5109716, RDS:RDS20099}.  

 Source tracking in low-frequency aperture arrays (LFAA) is done solely by electronic scanning since LFAA do not have moving parts. As a result, other than at zenith, the beam from each of the dual polarized elements is generally different due to ``projection'' (or element foreshortening) effects~\cite{Ord2010}. This difference leads to strong instrumental polarization which has been identified as a challenge in LFAA high precision polarimetry~\cite{2000prat.conf..323H, 2000SPIE.4015..353H}. Since a large amount of astrophysical information is encoded in the full Stokes parameters of the radiation field, it is critical that the polarization performance and calibration of LFAA is well understood~\cite{RDS:RDS5907}. 

\begin{figure}[b]
	\begin{center}
	\includegraphics[width=3.25in]{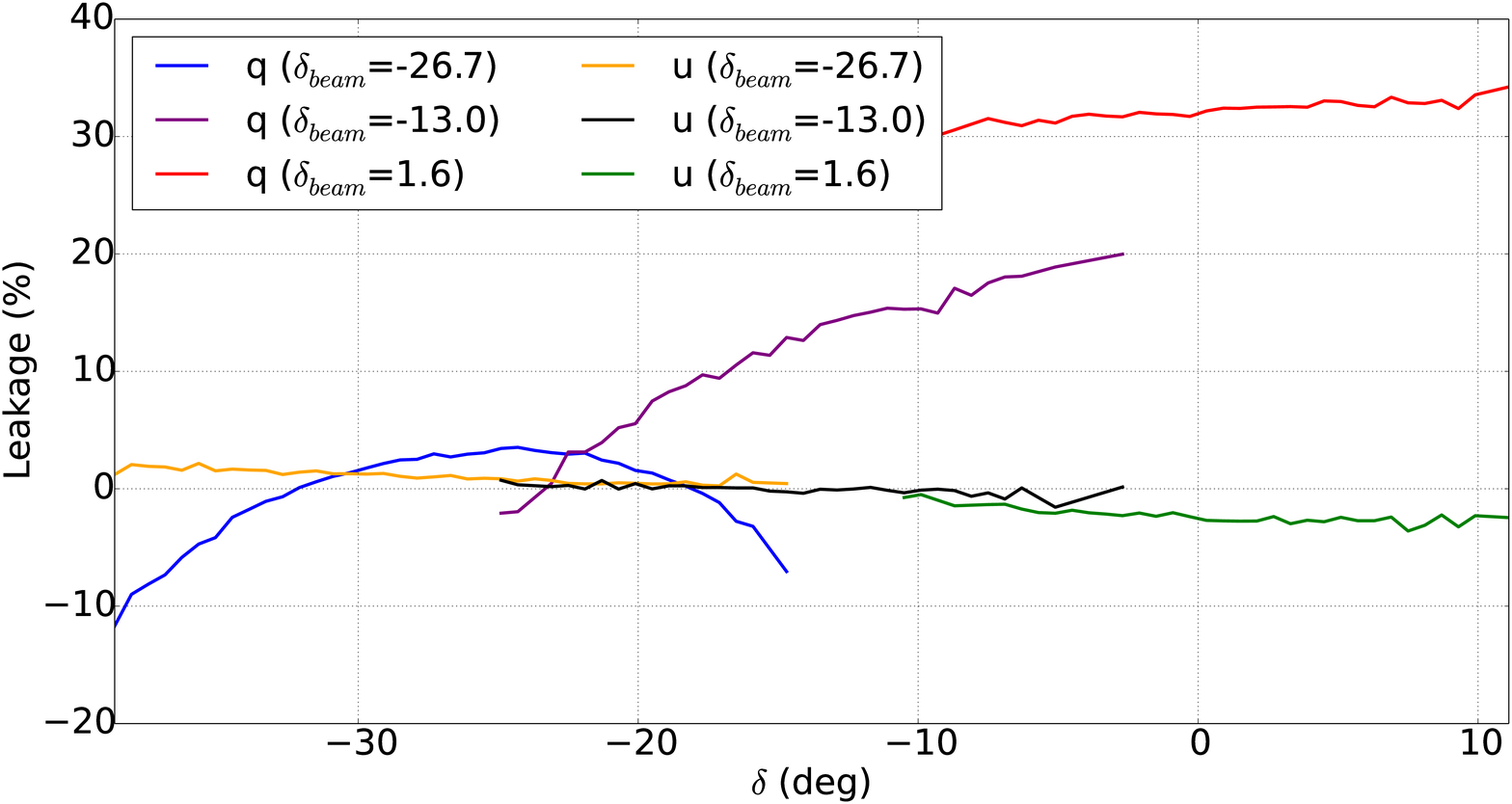} 
	\end{center}
	\caption{Median percentage polarization leakage (Q/I and U/I) measured for bright radio sources at various declinations ($\delta$) within $\pm0.1$ hour angle  with three different beam-former settings for scan angles along the meridian. The scan angles in Zenith angles are $Za=0,14,28^{\circ}$ where $Za=\delta-26.7^{\circ}$ (The Latitude at the MRO is 26.7~S). This 200-230 MHz (centered at approximately 216~MHz) data was calibrated with source PMN J0444-2809 (1.5 degrees south off zenith or $\delta=-28.2^{\circ}$).}
	\label{fig:GLEAM216}
\end{figure}

\begin{figure}[b]
	\begin{center}
	\includegraphics[width=3.25in]{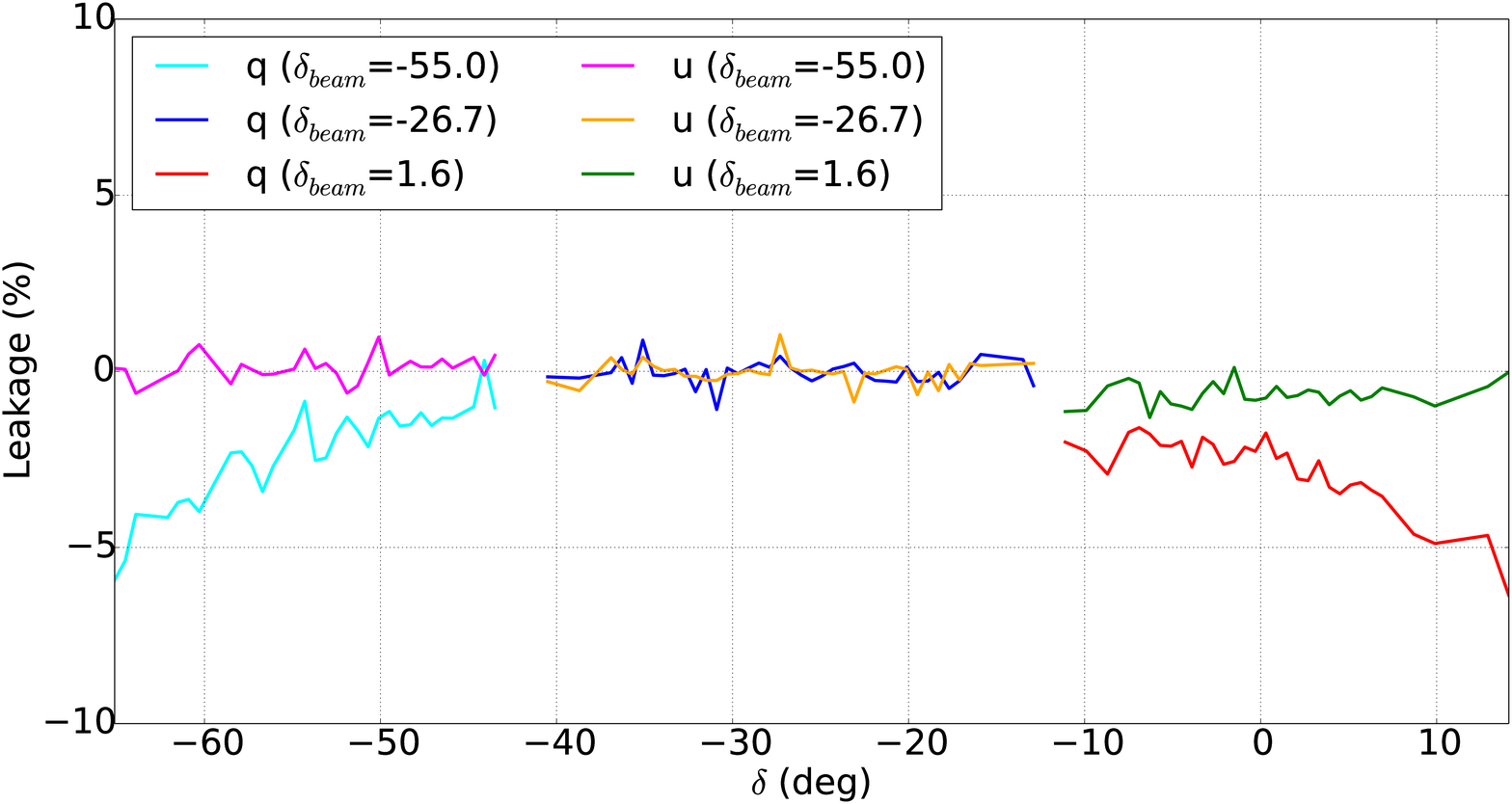}	
	\end{center}
	\caption{Median percentage polarization leakage (Q/I and U/I) measured for bright radio sources at various declinations ($\delta$) within $\pm0.1$ hour angle with three different beam-former settings for scan angles ($Za=-28,0,28^{\circ}$) along the meridian. This 140-170 MHz (centered at approximately 155~MHz) was calibrated with source PMN J0444-2809.}
	\label{fig:GLEAM155}
	\end{figure}

Strong residual instrumental polarization effects have indeed been seen in the MWA. A long-standing issue here is that calibration gain values for an MWA pointing angle do not transfer to other pointing angles~\cite{Ord2010}. Attempts to do so manifest most prominently as instrumental leakage in Stokes Q caused by ``X'' and ``Y'' beam patterns being very different from that predicted by simple pattern multiplication model. Fig.~\ref{fig:GLEAM216} demonstrates its severity, particularly at high frequencies (e.g., $>200$~MHz) where $Q/I$ for an unpolarized source may be as high as 20 to 30\% or more. At lower frequencies around 150~MHz, this error is of the order of a few percent (Fig.~\ref{fig:GLEAM155}). 

In this paper, we offer an explanation for these artifacts from a phased array theory perspective. The array is considered in transmission which is well known to be equal to the receive behavior using the principle of reciprocity. \textcolor{black}{It should be pointed out that the phased array theory we are about to describe are well known in the antenna engineering community. In this work, however, we report evaluation of the results in units relevant to astronomical polarimetry which, to our knowledge, is less common}. \textcolor{black}{The rest of this paper is organized as follows.} Section~\ref{sec:review} provides a simple review of astronomical polarimetry and clarifies the meanings of ``calibration'' and Stokes leakage in this context. Section~\ref{sec:model} generalizes the concept of pattern multiplication to include mutual coupling and discusses its significance to Q leakage. Section~\ref{sec:refine} addresses the shortcomings of the foregoing model and describes more rigorous models based on embedded element pattern. The performance of the most rigorous model (``full'' embedded element) is tested by simulating MWA observations reported in Figs.~\ref{fig:GLEAM216} and~\ref{fig:GLEAM155} and comparing the results to the measurement data. Next, we estimate anticipated improvement in Stokes leakage over the current MWA model afforded by the second most rigorous model (``average'' element pattern). Finally, lessons learned are summarized in Section~\ref{sec:concl}.

\section{Brief Review of Polarimetry}
\label{sec:review}
We begin by introducing the framework and key assumptions which we will follow in the rest of this paper. Our approach is electromagnetic and phased array theory based. Consequently, we take for granted the finer points of synthesis imaging such as electronic gain calibration~\cite{taylor_synthesis_1999} and the uniqueness of such solutions~\cite{2000prat.conf..323H, 2000SPIE.4015..353H}. Although admittedly far from being trivial for a real LFAA telescope, this presents no complication in our current approach as the ``tile'' can be excited with amplitude and phase under our control. In other words, our starting point is an array with equalized ``electronic gain.''  

Further, we assume identical time-independent MWA tiles for each baseline such that the ``apparent'' sky may be obtained by 2D inverse Fourier Transform of the visibilities~\cite{Smirnov:2011vp}. Hence, we are working in the ``image'' plane where the apparent sky in a given direction is given by~\cite{Smirnov:2011vp}
\begin{linenomath*}
\begin{eqnarray}
\mathbf{B}_{app}=\mathbf{J}\mathbf{B}\mathbf{J}^{H}
\label{eqn:V}
\end{eqnarray}
\end{linenomath*}
where $\mathbf{B}=\left\langle \mathbf{e}\mathbf{e}^H\right\rangle$ is the sky brightness matrix and $\mathbf{e}=(e_{\theta},e_{\phi})^{T}$. The subscripts $_{\theta}$ and $_{\phi}$ refer to the spherical coordinate bases (far-field unit) vectors~\cite{6420896, Sutinjo_TAP_TBP2014}. $\mathbf{J}$ is the presumed ``true'' Jones matrix that describes direction dependent effects for the tile. 
\begin{linenomath*}
\begin{eqnarray}
\mathbf{J}=
\left( \begin{array}{cc}
J_{x\theta} & J_{x\phi} \\
J_{y\theta} & J_{y\phi} 
\end{array} \right)
\label{eqn:J}
\end{eqnarray}
\end{linenomath*}
\noindent
where the subscripts $_{x.}$ and $_{y.}$ refer to measurement bases as per the ``X'' (E-W) and ``Y'' (N-S) bow-ties, respectively.

To obtain an estimate of the sky brightness matrix, $\tilde{\mathbf{B}}$, from the apparent sky, we use a Jones matrix model for the tile 
\begin{linenomath*}
\begin{eqnarray}
\tilde{\mathbf{B}}=\tilde{\mathbf{J}}^{-1}\mathbf{B}_{app}(\tilde{\mathbf{J}}^{H})^{-1}
\label{eqn:tildeB}
\end{eqnarray}
\end{linenomath*}
where $\tilde{\mathbf{J}}$ is our best estimate of the true Jones matrix for the tile. For simplicity, we further assume that the sources that drift through the array beam are unpolarized such that $\mathbf{B}=\mathbf{I}/2$ \textcolor{black}{(where $\mathbf{I}$ is the identity matrix)} and
\begin{linenomath*}
\begin{eqnarray}
\tilde{\mathbf{B}}=\frac{1}{2}\tilde{\mathbf{J}}^{-1}\mathbf{J}\mathbf{J}^{H}(\tilde{\mathbf{J}}^{H})^{-1}
\label{eqn:calibration_unpol}
\end{eqnarray} 
\end{linenomath*}
where $\tilde{\mathbf{B}}$ is our estimate of the ``true'' sky after calibration. To examine the polarization properties of the ``true'' sky estimate, we calculate Stokes parameters based on the elements of $\tilde{\mathbf{B}}$
\begin{linenomath*}
\begin{eqnarray}
\tilde{I}&=&\tilde{B}_{1,1}+\tilde{B}_{2,2} \nonumber \\
\tilde{Q}&=&\tilde{B}_{1,1}-\tilde{B}_{2,2}\nonumber \\
\tilde{U}&=&\tilde{B}_{1,2}+\tilde{B}_{2,1}\nonumber \\
\tilde{V}&=&j(\tilde{B}_{1,2}-\tilde{B}_{2,1})
\label{eqn:stokes}
\end{eqnarray} 
\end{linenomath*}
We further assume that the sources are unpolarized such that $\tilde{Q}$, $\tilde{U}$, and $\tilde{V}$ in (\ref{eqn:stokes}) are the residual instrumental Stokes leakages after calibration. The amount of leakage relative to $\tilde{I}$ is of primary interest, hence the results will be presented as normalized to $\tilde{I}$. Note that as a consequence of this normalization, \emph{scalar} multipliers to the Jones matrices become immaterial as they do not affect the results. For simplicity, the $\tilde{.}$ sign in reference to the Stokes leakages will be suppressed henceforth.

In the next two Sections, we will present a few examples involving assumed ``real'' Jones matrix, $\mathbf{J}$, and the estimate thereof, $\tilde{\mathbf{J}}$ to establish the following:
\begin{itemize}
\item Sec.~\ref{sec:model}:  $\mathbf{J}$ includes mutual coupling, $\tilde{\mathbf{J}}$ ignores mutual coupling. How significant is mutual coupling in our problem?
\item Sec.~\ref{sec:full}:  $\mathbf{J}$ is the most rigorous model (``full'' embedded element pattern), $\tilde{\mathbf{J}}$ is based on Hertzian dipole and ignores mutual coupling (current MWA practice). How reliable is our model? How well does it replicate the observation?
\item Sec.~\ref{sec:avg}:  $\mathbf{J}$ is based on the most rigorous ``full'' embedded element pattern, $\tilde{\mathbf{J}}$ is the second best model based on ``average'' embedded element pattern. What can we do to improve the situation and by how much?
\end{itemize}

\section{Simple Models}
\label{sec:model}
\subsection{Current Approach}
\label{sec:current}
The importance of modeling ``projection'' effects has been recognized in the MWA community. The current approach is purely geometrical which is derived from the dot products of incoming electric field vectors and $\hat{x}$ or $\hat{y}$ representing the bow-ties~\cite{Ord2010}. This is equivalent to modeling the elements as Hertzian dipoles. 

\begin{figure}[htb]
	\begin{center}
	{\includegraphics[width=2.5in]{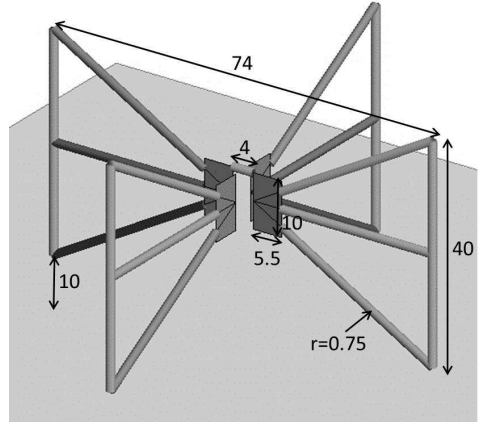}}
	\end{center}
\caption{MWA bow-ties as simulated in a commercial Method-of-Moment software FEKO. Dimensions are in cm. The distance to the perfect electric ground plane is 10~cm and the wire radius is 0.75~cm.}
\label{fig:MWA_BT}
\end{figure}

As a convention, we distinguish telescope pointing denoted by  superscript $.^{Az,Za}$ from direction dependent effects represented by spherical coordinates $(\theta,\phi)$ (where $\theta$ is zenith angle and $\phi=90^{\circ}-$azimuth is measured clockwise from $+x$ axis (E) looking down on the MWA tile, see Tab.~\ref{tab:numbering})
\begin{linenomath*}
\begin{eqnarray}
\tilde{\mathbf{J}}^{Az,Za}(\theta,\phi)=f^{Az,Za}(\theta,\phi)g(\theta)\mathbf{J}_{e}(\theta,\phi)
\label{eqn:Jdipole}
\end{eqnarray}
\end{linenomath*}
\noindent
where 
\begin{linenomath*}
\begin{eqnarray}
f^{Az,Za}&=&\frac{1}{16}\sum_{n=1}^{16}v^{Az,Za}_{n}e^{j(k_{x}x_{n}+k_{y}y_{n})} \label{eqn:f_ideal}\\
g(\theta)&=&\sin(2\pi\frac{h}{\lambda}\cos\theta)
\label{eqn:g_ideal}
\end{eqnarray}
\end{linenomath*}
where $f^{Az,Za}$ is the \emph{scalar} array factor; $k_{x}=(2\pi/\lambda)\sin\theta \cos\phi$ and $k_{y}=(2\pi/\lambda)\sin\theta \sin\phi$; $x_{n}, y_{n}$ are the feed point positions; $v^{Az,Za}_{n}$ is the excitation voltage based on the delay setting for a given pointing; $g(\theta)$)  is a \emph{scalar} that accounts for metallic ground plane effects where $h$ is the height above ground; and $\mathbf{J}_{e}(\theta,\phi)$ in this case is an ideal Jones matrix of a Hertzian dipole~\cite{Cheng_ch11,Balanis_1989_app}:
\begin{linenomath*}
\begin{eqnarray}
\mathbf{J}_{e}(\theta,\phi)=
\left( \begin{array}{cc}
\cos\theta\cos\phi & -\sin\phi  \\
\cos\theta\sin\phi  & \cos\phi
\end{array} \right)
\label{eqn:Je}
\end{eqnarray}
\end{linenomath*}
As seen earlier in Figs.~\ref{fig:GLEAM216} and~\ref{fig:GLEAM155}, this model seems to work well at the low frequencies but poorly at high frequencies, an observation consistent with our qualitative understanding of dipoles. Given the bow-tie dimensions in Fig.~\ref{fig:MWA_BT}, we note that the half-wavelength frequencies for the bow-tie length (74~cm) and diagonal (84~cm) occur at 203 and 179 MHz, respectively. Hence, below these frequencies the bow-ties are more or less short with respect to wavelength but not at higher frequencies. This has led to a follow-up effort where the bow-ties are represented as dipoles with sinusoidal current distribution and physical lengths~\cite{McKinley_2012}. This model, however, has not moved beyond the proposal stage to implementation. 

In the context of the simplest model discussed so far, we pose this question: is it realistic to expect a satisfactory solution based only on improving or perfecting $\mathbf{J}_{e}(\theta,\phi)$ alone? It seems not. Our inability to transfer calibration solutions between telescope pointing angles, especially as evidenced by a large discontinuity in Q leakage in Fig.~\ref{fig:GLEAM216} between $Za=14$~and~$28^{\circ}$, suggests that there exists instrumental imbalance between the ``X'' and ``Y'' that is unique for every pointing which is unaccounted for. We hypothesize that port currents present at the MWA bow-ties are in general not equal in amplitudes and deviate from the imposed voltages in (\ref{eqn:f_ideal}). For every pointing angle, there is a corresponding set of 32 port currents which do not in general ``transfer'' to other pointing angles. A simple mutual coupling model that describes this is presented next.

\subsection{Mutual Coupling Model}
\label{sec:mutual}
Assuming that the element patterns in the tile are identical such that pattern multiplication applies~\cite{Kraus_1988_ch4}, a salient feature of mutual coupling is the introduction of a diagonal array factor matrix, as opposed to a scalar (\ref{eqn:f_ideal}). Each entry in the diagonal matrix describes a distinct array factor for the ``X'' and ``Y'' elements, respectively as described in (\ref{eqn:Jmtl}) and (\ref{eqn:Axy}). 
\begin{linenomath*}
\begin{eqnarray}
\tilde{\mathbf{J}}^{Az,Za}(\theta,\phi)=\mathbf{A}_{xy}^{Az,Za}(\theta,\phi)g(\theta)\mathbf{J}_{e}(\theta,\phi) 
\label{eqn:Jmtl}
\end{eqnarray}
\end{linenomath*}
\noindent
where 
\begin{linenomath*}
\begin{eqnarray}
\mathbf{A}_{xy}^{Az,Za}(\theta,\phi)=
\left( \begin{array}{cc}
f_{x}^{Az,Za}(\theta,\phi) & 0\\
0 & f_{y}^{Az,Za}(\theta,\phi) 
\end{array} \right)
\label{eqn:Axy}
\end{eqnarray}
\end{linenomath*}
\noindent
and the array factors are
\begin{linenomath*}
\begin{eqnarray}
f^{Az,Za}_{y}=\frac{1}{16}\sum_{n=1}^{16}i^{Az,Za}_{n}e^{j(k_{x}x_{n}+k_{y}y_{n})} \nonumber \\
f^{Az,Za}_{x}=\frac{1}{16}\sum_{n=17}^{32}i^{Az,Za}_{n}e^{j(k_{x}x_{n}+k_{y}y_{n})}
\label{eqn:fxy}
\end{eqnarray}
\end{linenomath*}
with numbering convention, ``Y'' (port numbers 1 to 16) and ``X'' (port numbers 17 to 32) as shown in Tab.~\ref{tab:numbering} and $i^{Az,Za}_{1\ldots32}$ are the port currents obtained from (\ref{eqn:vZi}) as explained in the next paragraph. \textcolor{black}{For an array located over a metallic ground plane such as the MWA, the port currents in (\ref{eqn:fxy}) are the solutions of (\ref{eqn:vZi}) where the $\mathbf{Z}$ matrix is computed for an array over a ground plane.} 

\begin{table} [htb]
	\centering
	\begin{tabular}{c c c c}
	 1/17 & 2/18 & 3/19 & 4/20 \\
	 + & + & + & + \\
	  & &  &  \\
	 5/21 & 6/22 & 7/23 & 8/24 \\
	 + & + & + & + \\
	 & &  &  \\
	 9/25 & 10/26 & 11/27 & 12/28 \\
	 + & + & + & + \\
	 & &  &  \\
	 13/29 & 14/30 & 15/31 & 16/32 \\
	 + & + & + & + \\
	 & &  &  \\
	\end{tabular}
		\caption{Numbering convention: ``Y'' (N-S)/ ``X'' (E-W) dipoles in every entry ($\uparrow$ N, $\rightarrow$ E). Elements 1~/~17 are at the NW corner and 16~/~32 are at the SE corner.}
	\label{tab:numbering}
\end{table}

\begin{figure}[htb]
	\begin{center}
	{\includegraphics[width=3.25in]{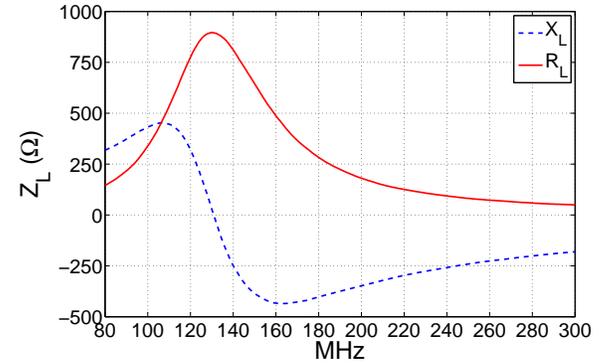}}
	\end{center}
\caption{Measured MWA LNA impedance}
\label{fig:MWA_LNA}
\end{figure}

At every frequency and pointing angle, the port currents, $\mathbf{i}=[i_{1}\ldots i_{16}, i_{17}\ldots i_{32}]^{T}$, and voltage delays  ,$\mathbf{v}=[v_{1}\ldots v_{16}, v_{17}\ldots v_{32}]^{T}$, are related via an impedance matrix which in the case of the MWA is 32$\times$32~\cite{Stutz_1998, Hirasawa_1992_ch2, Hansen_1998_ch7} corresponding to 16~``X'' and 16~``Y'' bow-ties.
\begin{linenomath*}
\begin{eqnarray}
\mathbf{i}=(\mathbf{Z}+\mathbf{Z}_{L})^{-1}\mathbf{v}
\label{eqn:vZi}
\end{eqnarray}
\end{linenomath*}
\noindent
where 
$\mathbf{v}$ is the known ``excitation'' voltage vector determined by delay settings, $\mathbf{i}$ is the unknown port current vector, the impedance parameter matrix $\mathbf{Z}$~\cite{Pozar_1998_ch4, Stutz_1998} describes the inter-port coupling and is given by 
\begin{linenomath*}
\begin{eqnarray}
\left( \begin{array}{ccc}
Z_{1,1} &\ldots &Z_{1,32}\\
\vdots &\ddots &\vdots\\
Z_{32,1} & \ldots &Z_{32,32}
\end{array} \right)
\label{eqn:Zpar}
\end{eqnarray}
\end{linenomath*}
Note that if mutual coupling is ignored, (\ref{eqn:Zpar}) becomes diagonal such that $\mathbf{v}\propto\mathbf{i}$. Consequently, port currents amplitudes are identical since the voltage amplitudes are uniform in the MWA. The load impedance matrix $\mathbf{Z}_{L}$ is a \emph{diagonal} matrix, the entries of which are the LNA input impedances seen at the corresponding ports. The measured impedance of an MWA LNA is reported in Fig.~\ref{fig:MWA_LNA}.

\begin{figure}[htb]
	\begin{center}
	{\includegraphics[width=3.25in]{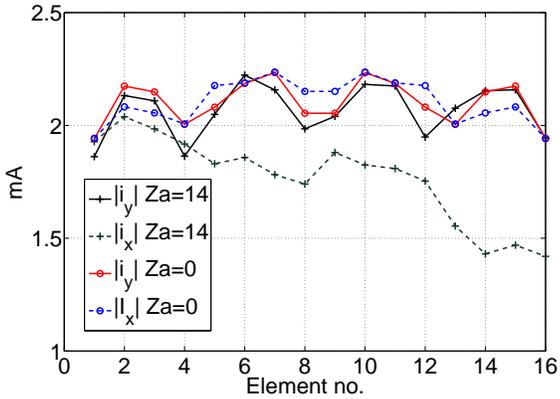}}
	\end{center}
\caption{Amplitude of currents at the ``X'' and ``Y'' ports at 216 MHz and $Az=0^{\circ},Za=0,14^{\circ}$ pointing angles for an MWA tile simulated in FEKO. The excitation voltage amplitude is 1~V at every port. The array is placed on a 5$\times$5~m perfect electric ground sheet on a soil model based on MRO sample with 2.5\% moisture.}
\label{fig:Ixy}
\end{figure}

\begin{figure}[htb]
	\begin{center}
	{\includegraphics[width=1.5in]{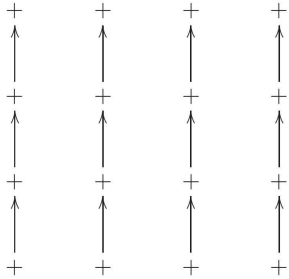}}
	\end{center}
\caption{Phase gradient when pointing the telescope on $Az=0^{\circ}$ plane for $0^{\circ}<Za<90^{\circ}$. The arrows are directed along the direction of increasing phase delay.}
\label{fig:phasegrad}
\end{figure}

This model now allows us examine our hypothesis. Fig.~\ref{fig:Ixy} reports current amplitudes at telescope pointing angles $Az=0^{\circ}$ and $Za=0,14^{\circ}$ at 216 MHz. Note that at $Za=0^{\circ}$, $|i_{x}|$ and $|i_{y}|$ occupy similar values. However, at $Za=14^{\circ}$, $|i_{x}|$ is lower than $|i_{y}|$ by approximately 10 to 20\%. Mutual coupling leads to the difference in the ``X'' and ``Y'' port current amplitudes even though the excitation voltage amplitudes are uniform at all ports. This effect is particularly pronounced on the principal planes (i.e, $\phi=0, 90, 180, 270^{\circ}$) as the interactions between ``X'' and ``Y'' bow-ties are asymmetric with respect to the phase delay gradient as illustrated in Fig.~\ref{fig:phasegrad}. The arrows there are drawn in the direction of increasing phase delay. Note that the phase delay gradient forces ``side-to-side'' interactions on the ``X'' bow-ties as opposed to ``end-to-end'' interactions on the ``Y'' bow-ties resulting in current deviation seen in Fig.~\ref{fig:Ixy}. Such asymmetry does not occur when pointing the telescope in the diagonal plane as the interactions are symmetric with respect to the ``X'' and ``Y'' bow-ties. 

\begin{figure}[htb]
	\begin{center}
	{\includegraphics[width=3.25in]{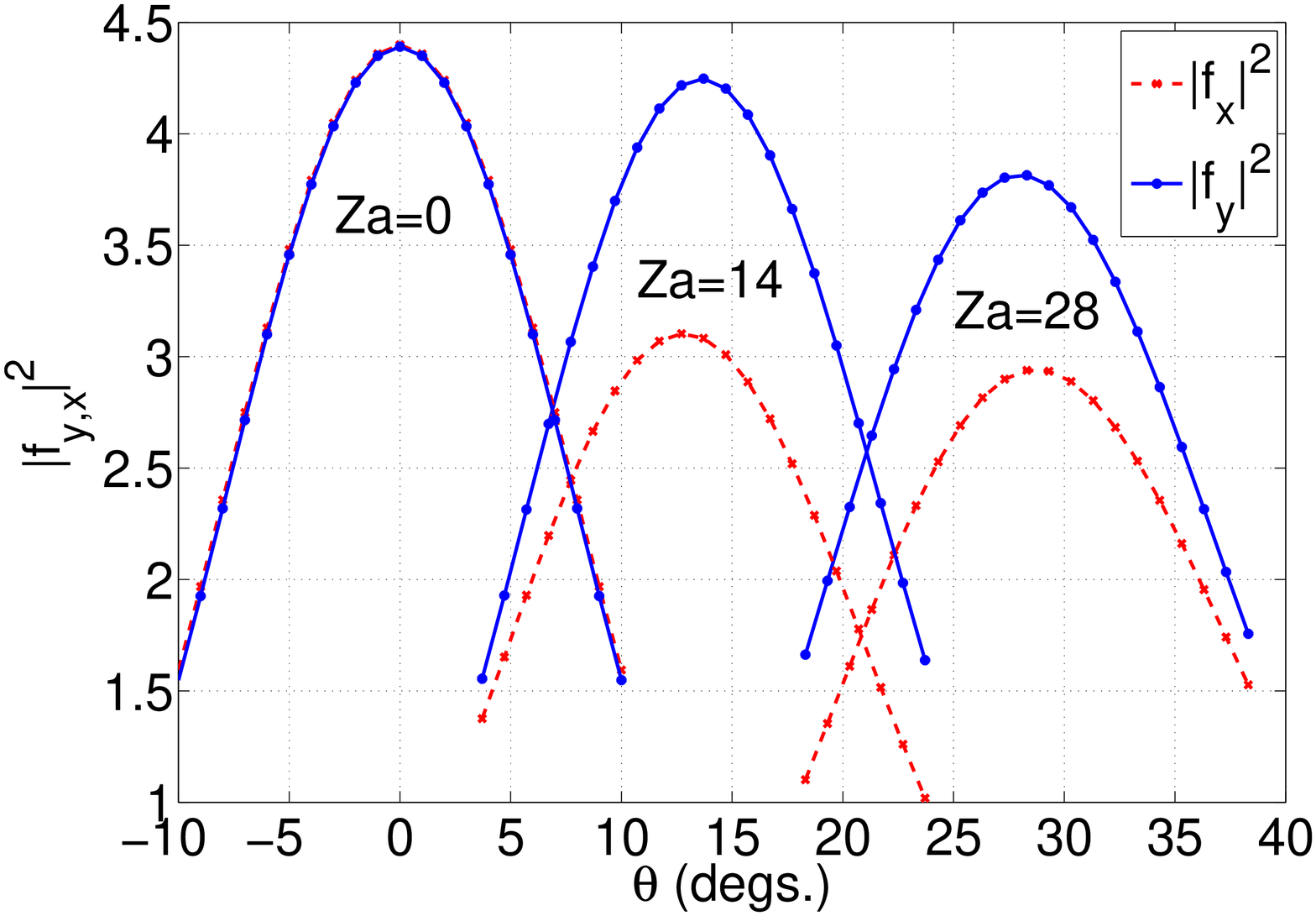}}
	\end{center}
\caption{Amplitude squared of the array factors for the ``X'' and ``Y'' bow-ties at 216 MHz and $Az=0^{\circ},Za=0,14,28^{\circ}$ pointing angles on $\phi=90^{\circ}$ plane for an MWA tile simulated in FEKO. The excitation voltage amplitude is 1~V at every port.}
\label{fig:fxy_216}
\end{figure}

As expected, the imbalances in the current amplitudes are reflected in the array factors for the ``X'' and ``Y'' bow-ties as described in (\ref{eqn:fxy}). Fig.~\ref{fig:fxy_216} reports $|f_{x}|^2$ and $|f_{y}|^2$ at 216 MHz for $Az=0^{\circ},Za=0,14,28^{\circ}$ on $\phi=90^{\circ}$ plane. The array factors are very similar only when the tile points at zenith. For $Za=14,28^{\circ}$ notable deviations in $|f_{x}|^2$ and $|f_{y}|^2$ are seen. Thus, for example, if the telescope is electronic gain calibrated at zenith (which is consistent with Fig.~\ref{fig:fxy_216}) using simple model in (\ref{eqn:Jdipole}), these solutions will not transfer to $Za=14,28^{\circ}$ as the deviations in $|f_{x}|^2$ and $|f_{y}|^2$ at those pointing angles cannot be accommodated by the simple model. Furthermore, given the definitions of Stokes Q leakage in (\ref{eqn:stokes}) and our model in (\ref{eqn:Jmtl}), we suspect that this imbalance in ``X'' and ``Y'' is a significant contributor to the Q leakage. This is quantified next. 

\subsection{Emergence of Q leakage}
\label{sec:Q leakage}
The appearance of Q leakage may be demonstrated by revisiting (\ref{eqn:tildeB}). Let the mutual coupling model in (\ref{eqn:Jmtl}) be taken as ``reality'' ($\mathbf{J}$) to be estimated by the simple model in (\ref{eqn:Jdipole}) that neglects mutual coupling. We further assume an optimistic scenario where the element pattern $\mathbf{J}_{e}(\theta,\phi)$ is fully known (i.e., identical in ``reality'' and the model). Here, the beam effects are completely canceled through calibration such that we are calculating Q leakage due solely to mutual coupling effects. 
\begin{linenomath*}
\begin{eqnarray}
\tilde{\mathbf{B}}&=&\frac{1}{2|f g|^{2}}\tilde{\mathbf{J}}_{e}^{-1}\mathbf{J}\mathbf{J}^{H} (\tilde{\mathbf{J}}_{e}^{H})^{-1} \nonumber \\
&=&\frac{1}{2|f g|^{2}}\mathbf{J}_{e}^{-1}\mathbf{A}_{xy}\mathbf{J}_{e}\mathbf{J}_{e}^{H}\mathbf{A}_{xy}^{H}(\mathbf{J}_{e}^{H})^{-1}
\label{eqn:tildeB_unpol}
\end{eqnarray}
\end{linenomath*}

\begin{figure}[htb]
	\begin{center}
	{\includegraphics[width=3.25in]{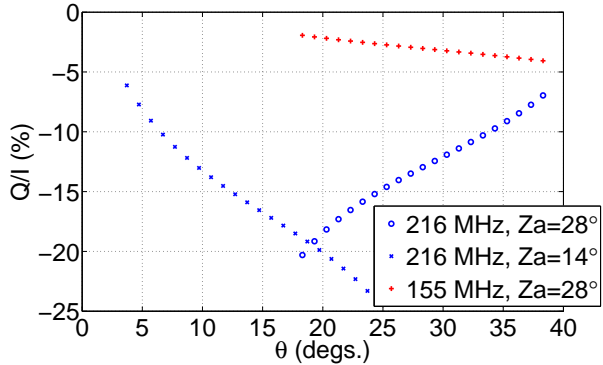}}
	\end{center}
\caption{$Q/I^{Az=0^{\circ},Za}(\theta,\phi=90^{\circ})$ at 155 MHz and $Za=28^{\circ}$, 216 MHz and $Za=14, 28^{\circ}$ for an MWA tile simulated in FEKO. The array is placed on a 5$\times$5~m perfect electric ground sheet on a soil model based on MRO sample with 2.5\% moisture~\cite{SP_soil}.}
\label{fig:QI_emil}
\end{figure}

Observation data shown in Figs.~\ref{fig:GLEAM216} and~\ref{fig:GLEAM155} were taken at three zenith angles and $Az=0^{\circ}$. Limiting further only to sources on the $\phi=90^{\circ}$ (meridional) plane and assuming the cross bow-ties have low raw cross-polarization components on the principal planes, one may approximate $\mathbf{J}_{e}(\theta,\phi=90^{\circ})$ as a diagonal matrix allowing it to commute with $\mathbf{A}_{xy}$. As a result, on this $\phi=90^{\circ}$ principal plane we are left with
\begin{linenomath*}
\begin{eqnarray}
\tilde{\mathbf{B}}^{Az=0^{\circ},Za}(\theta,\phi=90^{\circ})&=&\frac{1}{2|f g|^{2}}\mathbf{A}_{xy}\mathbf{A}_{xy}^{H} \nonumber \\
&=&\frac{1}{2|f g|^{2}}\left( \begin{array}{cc}
|f_{x}|^{2} & 0\\
0 & |f_{y}|^{2} 
\end{array} \right)
\label{eqn:tildeB_simple}
\end{eqnarray}
\end{linenomath*}
We see that imbalance in $|f_{x}|^2$ and $|f_{y}|^2$ registers as Q leakage after calibration if port coupling is not accounted for. Normalizing to Stokes I, we can write
\begin{linenomath*}
\begin{eqnarray}
\frac{Q}{I}^{Az=0^{\circ},Za}(\theta,\phi=90^{\circ})=\frac{|f_{x}|^{2}-|f_{y}|^{2}}{|f_{x}|^{2}+|f_{y}|^{2}}
\label{eqn:QI}
\end{eqnarray}
\end{linenomath*}
To compare to Figs.~\ref{fig:GLEAM216} and~\ref{fig:GLEAM155}, we report numerical results in Fig.~\ref{fig:QI_emil} of $Q/I$ per (\ref{eqn:QI}) at 155 and 216~MHz (center frequencies of the observations). $Q/I$ at 155~MHz appears consistent with observation in that it is increasingly negative away from zenith to a few percent leakage. At 216~MHz, $Q/I$ could reach negative 20\% range due to significant imbalance in $|f_{x}|^2$ and $|f_{y}|^2$ as seen in Fig.~\ref{fig:fxy_216}. Note that at 216~MHz, as $|f_{x}|^2<|f_{y}|^2$, the sign of $Q/I$ is opposite to the observation data which suggests that the assumption of identical element pattern in (\ref{eqn:tildeB_unpol}) is overly simplistic at this frequency. This will be dealt with in the next Section where the interplay between element pattern and mutual coupling is accounted for more rigorously. Nevertheless, it is important to note that one can expect up to approximately 20\% residual Q leakage at high frequency if mutual coupling is ignored.

\section{Model Refinement}
\label{sec:refine}
We now introduce more accurate models. The most rigorous full embedded element pattern model is described and subsequently taken as reality. Two reduced models are then used as Jones matrix estimates: simple model in (\ref{eqn:Jdipole}) and the average embedded element pattern. As discussed in Sec.~\ref{sec:current}, the Hertzian dipole model reflects current MWA practice. The average embedded element pattern model is less rigorous than the full embedded element pattern; however, \textcolor{black}{since new observation data with the new models is not yet available, this calculation} provides the best available estimate of the level of improvement expected from an upgraded model.

\subsection{Full Embedded Element Pattern}
\label{sec:full}
The full embedded element pattern rigorously accounts for mutual coupling and variations in the element pattern in a finite array~\textcolor{black}{\cite{273305, 310010, Hansen_1998_ch7, 1016399}}. It does not rely on pattern multiplication. Hence, it is appropriate here to take it as simulated reality. For the MWA, this model requires calculations of 32 embedded element patterns (16 for each polarization) at each frequency. 

The Jones matrix for a given ($Az,Za$) pointing is given by~\cite{1016399}:
\begin{linenomath*}
\begin{eqnarray}
\mathbf{J}^{Az,Za}(\theta,\phi)=
\frac{1}{16}\sum_{n=1}^{16}
\left( \begin{array}{cc}
i_{n+16}^{Az,Za} & 0\\
0 & i_{n}^{Az,Za}
\end{array} \right)
\mathbf{J}_{n}^{OC}(\theta,\phi)
\label{eqn:full}
\end{eqnarray}
\end{linenomath*}
\noindent
where $i_{1\ldots16}^{Az,Za}$ and $i_{17\ldots32}^{Az,Za}$ are the port currents for ``Y'' and ``X'' bow-ties as explained in (\ref{eqn:vZi}). $\mathbf{J}_{n}^{OC}(\theta,\phi)$ is the ``open circuit''~\cite{1016399} embedded Jones matrix for antenna \emph{pair} $n$ (involving index $n$ ``Y'' and index $n+16$ ``X'' dipoles):
\begin{linenomath*}
\begin{eqnarray}
\mathbf{J}_{n}^{OC}=
\left( \begin{array}{cc}
J^{x\theta}_{n+16} & J^{x\phi}_{n+16}\\
J^{y\theta}_{n} & J^{y\phi}_{n}
\end{array} \right)
\label{eqn:Joc}
\end{eqnarray}
\end{linenomath*}
where the superscripts $^{\theta}$ and $^{\phi}$ refer to spherical coordinate sky bases. Each row in (\ref{eqn:Joc}) is obtained by open circuiting every other port and exciting dipole $n$ or $n+16$ with a unit current and zero phase~\cite{1016399}. In this instance, $\mathbf{J}_{n}^{OC}$ is obtained using a FEKO model of an MWA tile on a perfect electric ground sheet placed on MRO soil as before.  

\begin{figure*}[htb]
	\begin{center}
	{\includegraphics[width=4.25in]{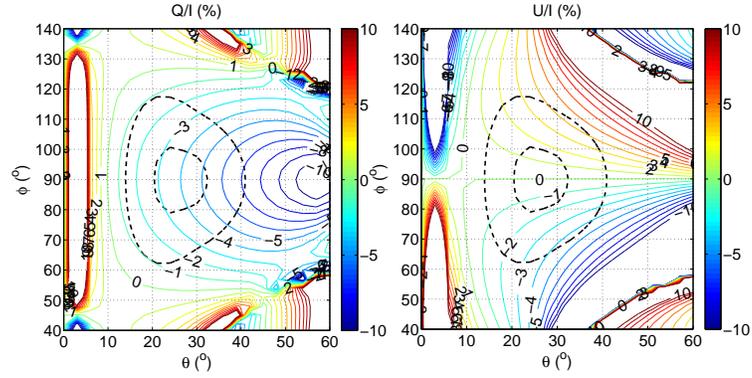}}
	\end{center}
\caption{Q and U Stokes leakages at 155~MHz for $Az=0^{o}, Za=28^{o}$  pointing after calibration with Hertzian dipole model with no mutual coupling. Normalized power gain contours of 0.5 and 0.9 are superimposed on contour plots with 1\% step per contour.}
\label{fig:155_28_short}
\end{figure*}

\begin{figure*}[htb]
	\begin{center}
	{\includegraphics[width=4.25in]{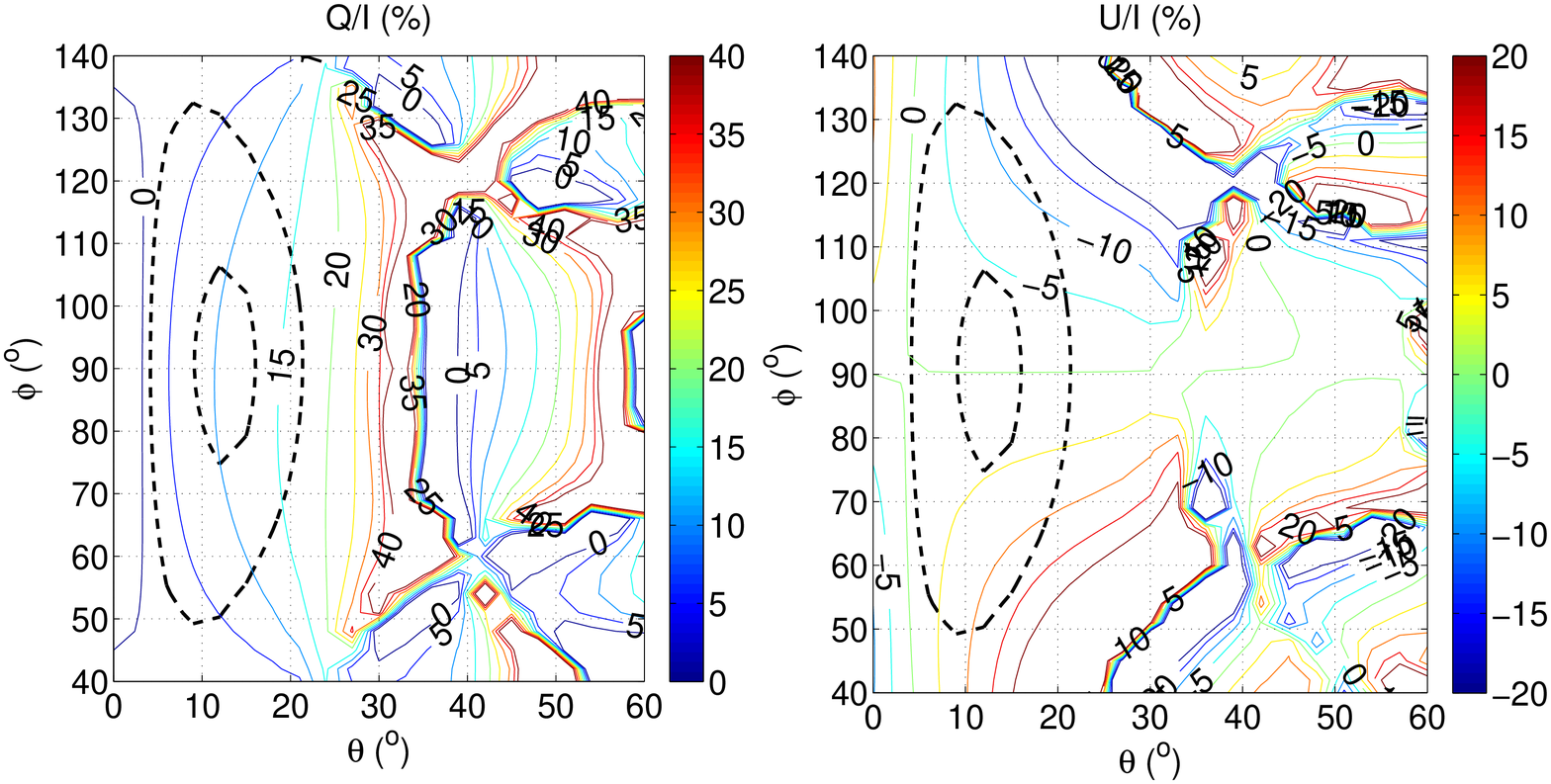}}
	\end{center}
\caption{Q and U Stokes leakages at 216~MHz for $Az=0^{o}, Za=14^{o}$  pointing after calibration with Hertzian dipole model with no mutual coupling. Normalized power gain contours of 0.5 and 0.9 are superimposed on contour plots with 5\% step per contour.}
\label{fig:216_14_short}
\end{figure*}

\begin{figure*}[htb]
	\begin{center}
	{\includegraphics[width=4.25in]{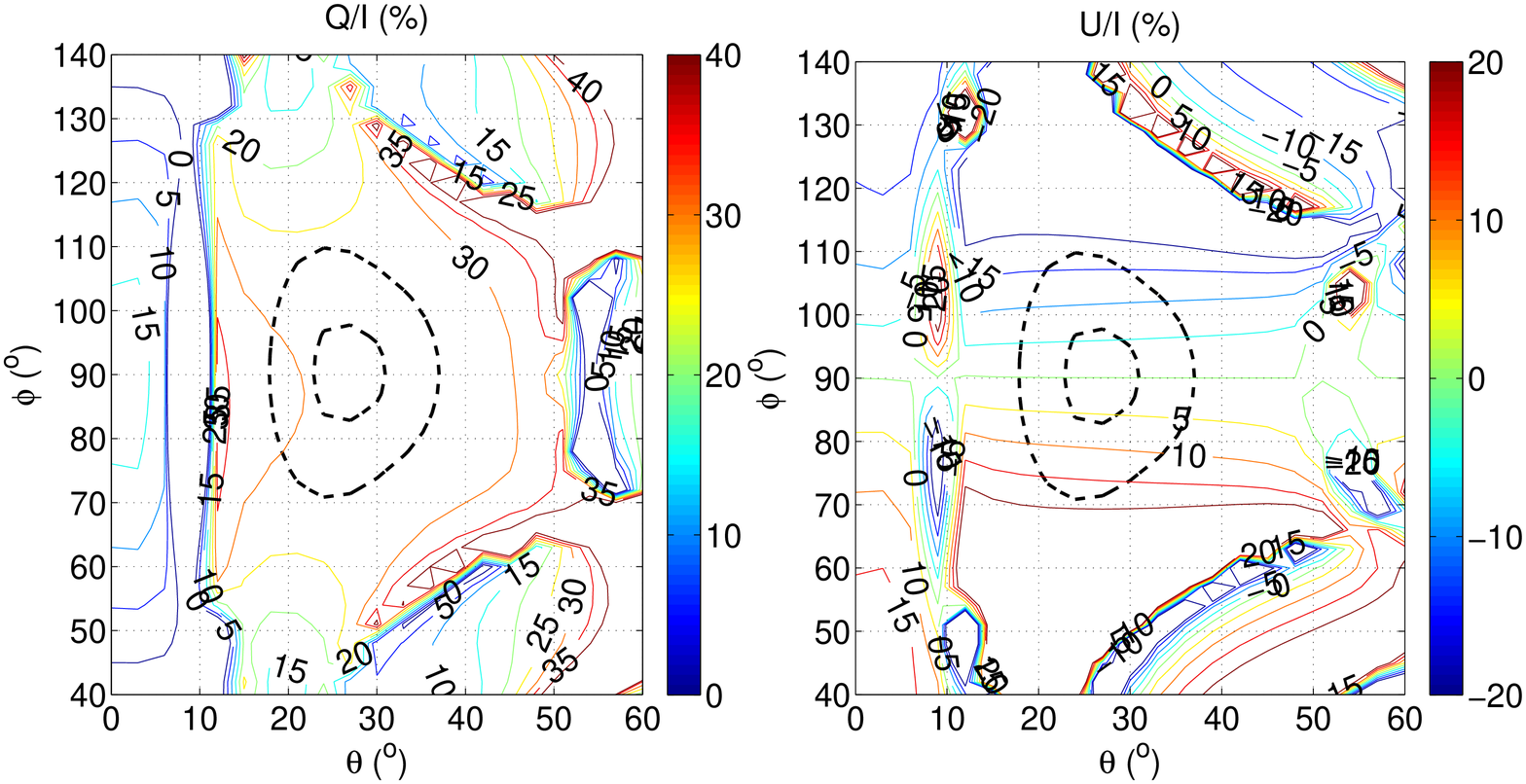}}
	\end{center}
\caption{Q and U Stokes leakages at 216~MHz for $Az=0^{o}, Za=28^{o}$  pointing after calibration with Hertzian dipole model with no mutual coupling. Normalized power gain contours of 0.5 and 0.9 are superimposed on contour plots with 5\% step per contour.}
\label{fig:216_28_short}
\end{figure*}

Calculated Stokes Q and U leakages at 155 and 216~MHz at  $Az=0, Za=14,28^{\circ}$ are reported in Figs.~\ref{fig:155_28_short},~\ref{fig:216_14_short}, and~\ref{fig:216_28_short}. Direct comparisons between observed and calculated $Q/I$ (on $Az=0^{\circ}$ plane) are reported in Figs.~\ref{fig:155_28_cut},~\ref{fig:216_14_cut} and~\ref{fig:216_28_cut}. The full model with simple Jones matrix estimate is indicated by legend ``Hertzian.'' Observed median $Q/I$ are reported in black dots. Note that the model replicates measured trends and values quite well. At 155~MHz and $Za=28^{\circ}$, $Q/I$ becomes more negative with increasing zenith angle (Fig.~\ref{fig:155_28_cut}). The difference between observation and calculation  is approximately 2\%. At 216~MHz and $Za=14^{\circ}$, $Q/I$ becomes more positive with zenith angle (Fig.~\ref{fig:216_14_cut}). Similarly, a difference of approximately 2\% is seen between calculation and observation. Finally, at 216~MHz and $Za=28^{\circ}$, the trends are again similar where $Q/I$ leakages are both very high and do not exhibit a steep slope with zenith angle (Fig.~\ref{fig:216_28_cut}). Here, calculated results are within roughly 5\% observed values. Calculated U leakages in Figs.~\ref{fig:155_28_short},~\ref{fig:216_14_short}, and~\ref{fig:216_28_short} indicate zero leakages on $Az=0^{\circ}$ plane. These are within a couple of percent of observed values in Figs.~\ref{fig:GLEAM216} and \ref{fig:GLEAM155}. These results provide evidence for the reliability of the full embedded element pattern.

\subsection{Average Embedded Element Pattern}
\label{sec:avg}
We turn our attention to the next best model based on ``average embedded element pattern.'' Again, the motivation here is to provide an estimate for the improvement in Stokes leakage expected from upgrading the Jones matrix estimate from simple Hertzian dipole to one based on embedded element pattern. The Jones matrix for the array is approximately~\cite{1016399}
\begin{linenomath*}
\begin{eqnarray}
\tilde{\mathbf{J}}^{Az,Za}(\theta,\phi)=
\left( \begin{array}{cc}
f_{x}^{Az,Za}(\theta,\phi) & 0\\
0 & f_{y}^{Az,Za}(\theta,\phi) 
\end{array} \right)
\mathbf{J}_{avg}^{OC}(\theta,\phi)
\label{eqn:Javg}
\end{eqnarray}
\end{linenomath*}
\noindent
where 
$\mathbf{J}_{avg}^{OC}(\theta,\phi)$ is the average embedded element pattern~\cite{273305} obtained by
\begin{linenomath*}
\begin{eqnarray}
\mathbf{J}_{avg}^{OC}=\frac{1}{16}\sum_{n=1}^{16}\mathbf{J}_{n}^{OC}e^{-j(k_{x}x_{n}+k_{y}y_{n})} 
\label{eqn:JavgOC}
\end{eqnarray}
\end{linenomath*}
where $\mathbf{J}_{n}^{OC}$ is obtained from (\ref{eqn:full}) and the exponential term compensates for the location of the dipole pair~$n$ such that the average is taken over the same phase reference point \emph{with respect to each active feed point}, for example, $\mathbf{J}_{1}^{OC}e^{-j(k_{x}x_{1}+k_{y}y_{1})}$ effectively moves the origin to dipole pair~1.

\begin{figure*}[htb]
	\begin{center}
	{\includegraphics[width=2.5in]{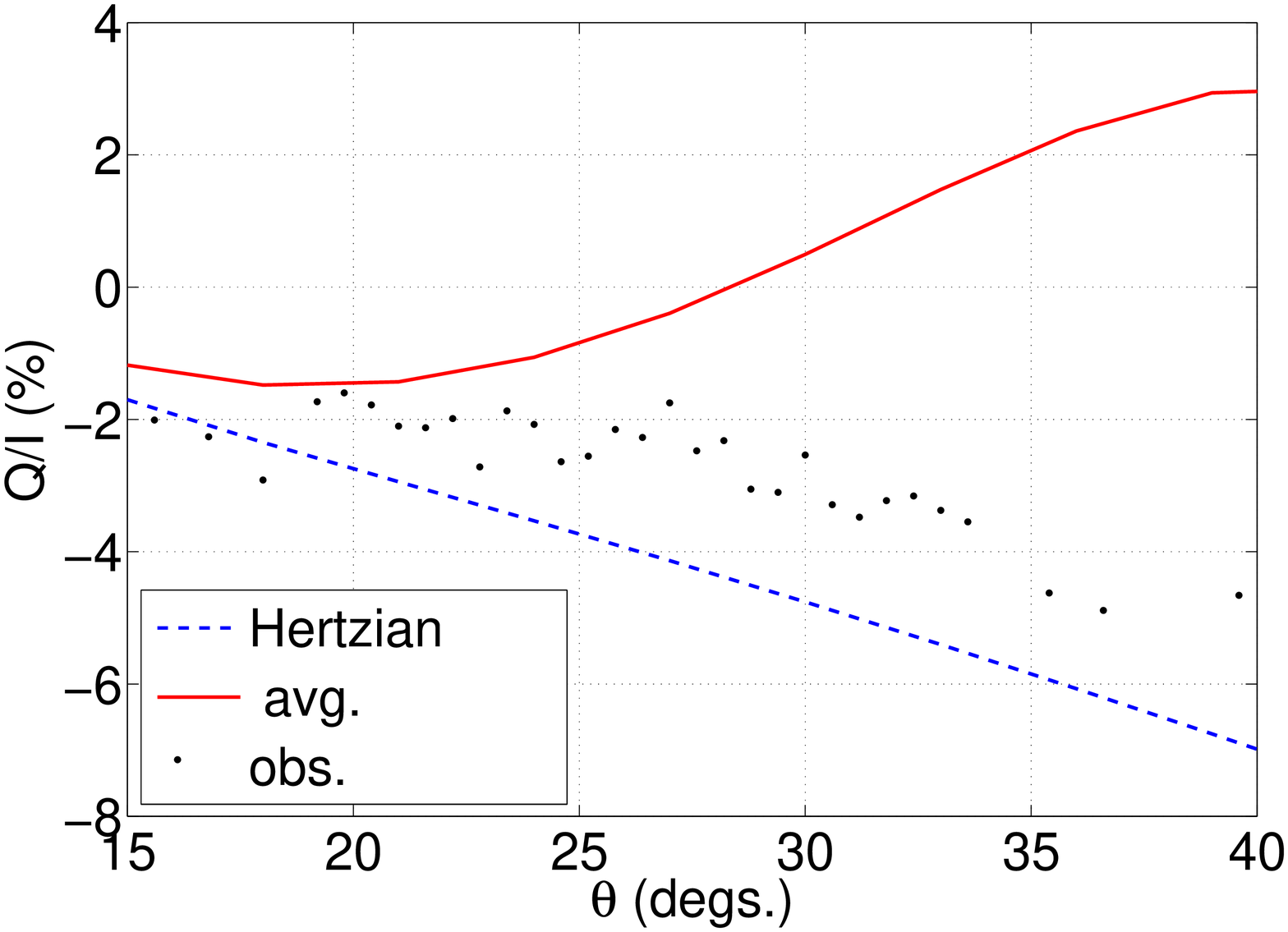}}
	\end{center}
\caption{$Q/I^{Az=0^{\circ},Za=28^{\circ}}(\theta,\phi=90^{\circ})$ at 155 MHz for an MWA tile simulated in FEKO. The full embedded element pattern shown in (\ref{eqn:full}) is assumed as ``reality.'' Two reduced models are used as Jones matrix estimates: the simple (``Hertzian'') model in (\ref{eqn:Jdipole}) and the average (``avg.'') embedded element pattern in (\ref{eqn:Javg}). Observed $Q/I$ with simple Jones matrix estimate as implemented in MWA are reported in black dots.}
\label{fig:155_28_cut}
\end{figure*}

\begin{figure*}[htb]
	\begin{center}
	{\includegraphics[width=2.5in]{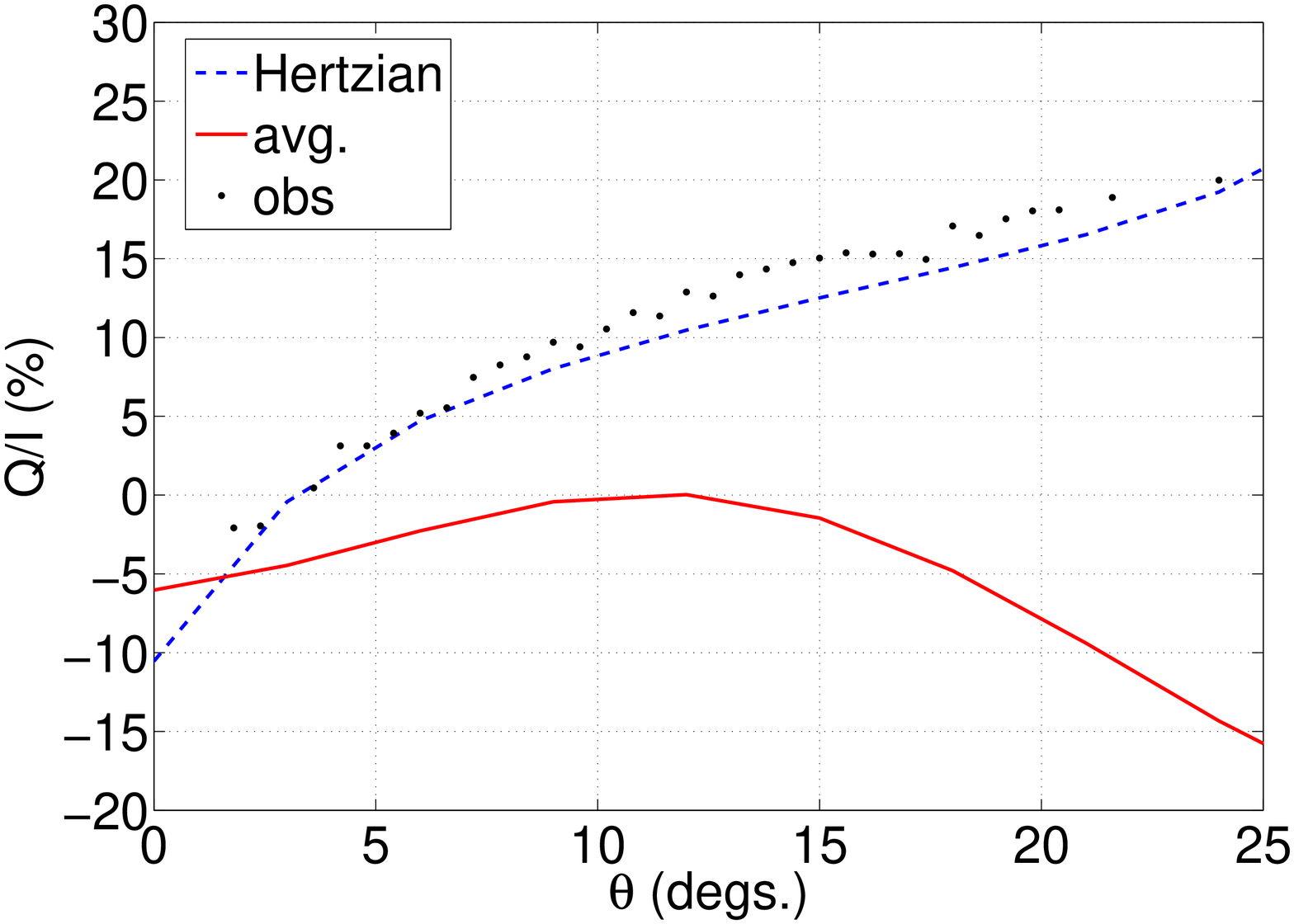}}
	\end{center}
\caption{$Q/I^{Az=0^{\circ},Za=14^{\circ}}(\theta,\phi=90^{\circ})$ at 216 MHz for an MWA tile simulated in FEKO. The full embedded element pattern shown in (\ref{eqn:full}) is assumed as ``reality.'' Two reduced models are used as Jones matrix estimates: the simple (``Hertzian'') model in (\ref{eqn:Jdipole}) and the average (``avg.'') embedded element pattern in (\ref{eqn:Javg}). Observed $Q/I$ with simple Jones matrix estimate as implemented in MWA are reported in black dots.}
\label{fig:216_14_cut}
\end{figure*}

\begin{figure*}[htb]
	\begin{center}
	{\includegraphics[width=2.5in]{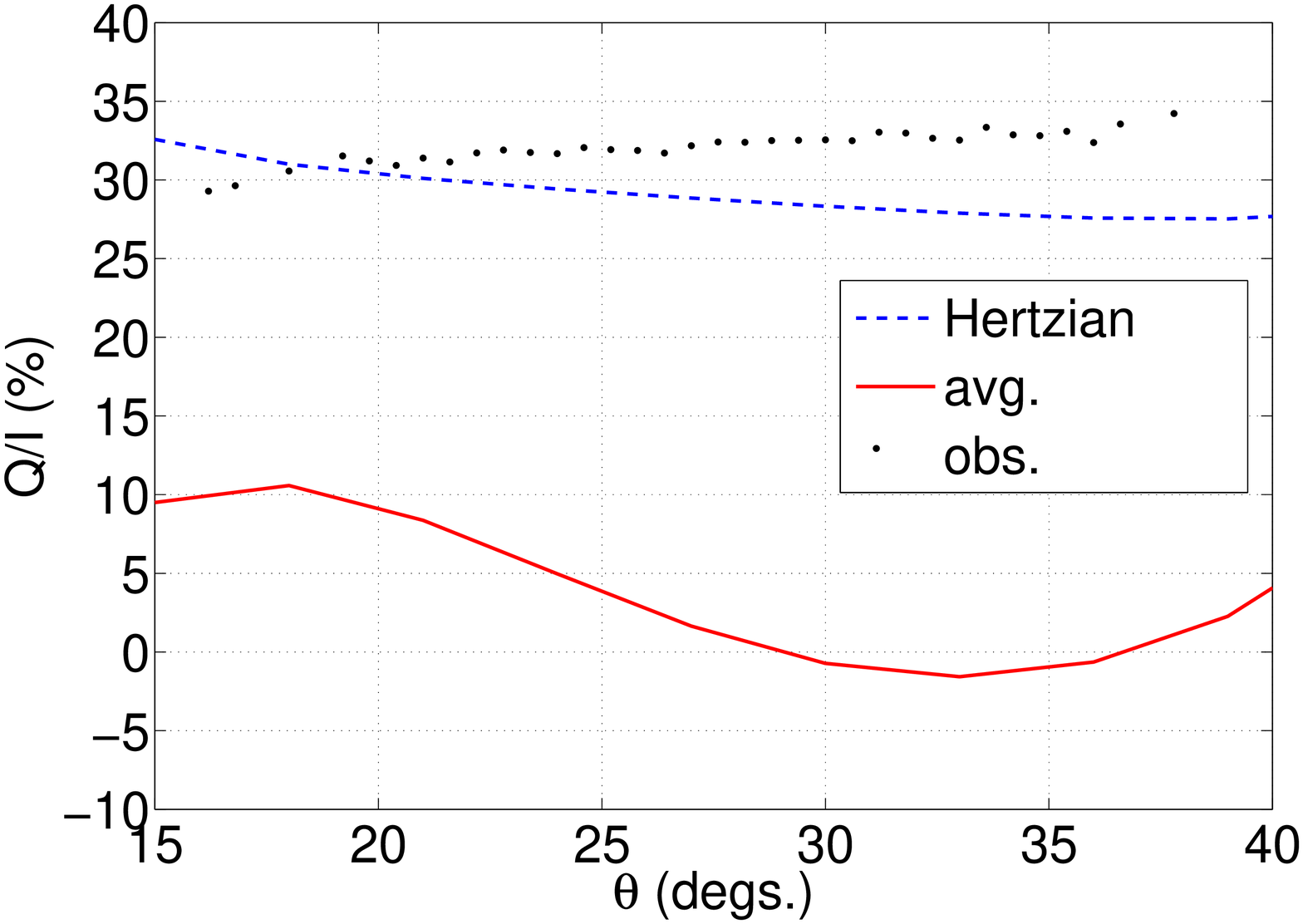}}
	\end{center}
\caption{$Q/I^{Az=0^{\circ},Za=28^{\circ}}(\theta,\phi=90^{\circ})$ at 216 MHz for an MWA tile simulated in FEKO. The full embedded element pattern shown in (\ref{eqn:full}) is assumed as ``reality.'' Two reduced models are used as Jones matrix estimates: the simple (``Hertzian'') model in (\ref{eqn:Jdipole}) and the average (``avg.'') embedded element pattern in (\ref{eqn:Javg}). Observed $Q/I$ with simple Jones matrix estimate as implemented in MWA are reported in black dots.}
\label{fig:216_28_cut}
\end{figure*}

\begin{figure*}[htb]
	\begin{center}
	{\includegraphics[width=4.25in]{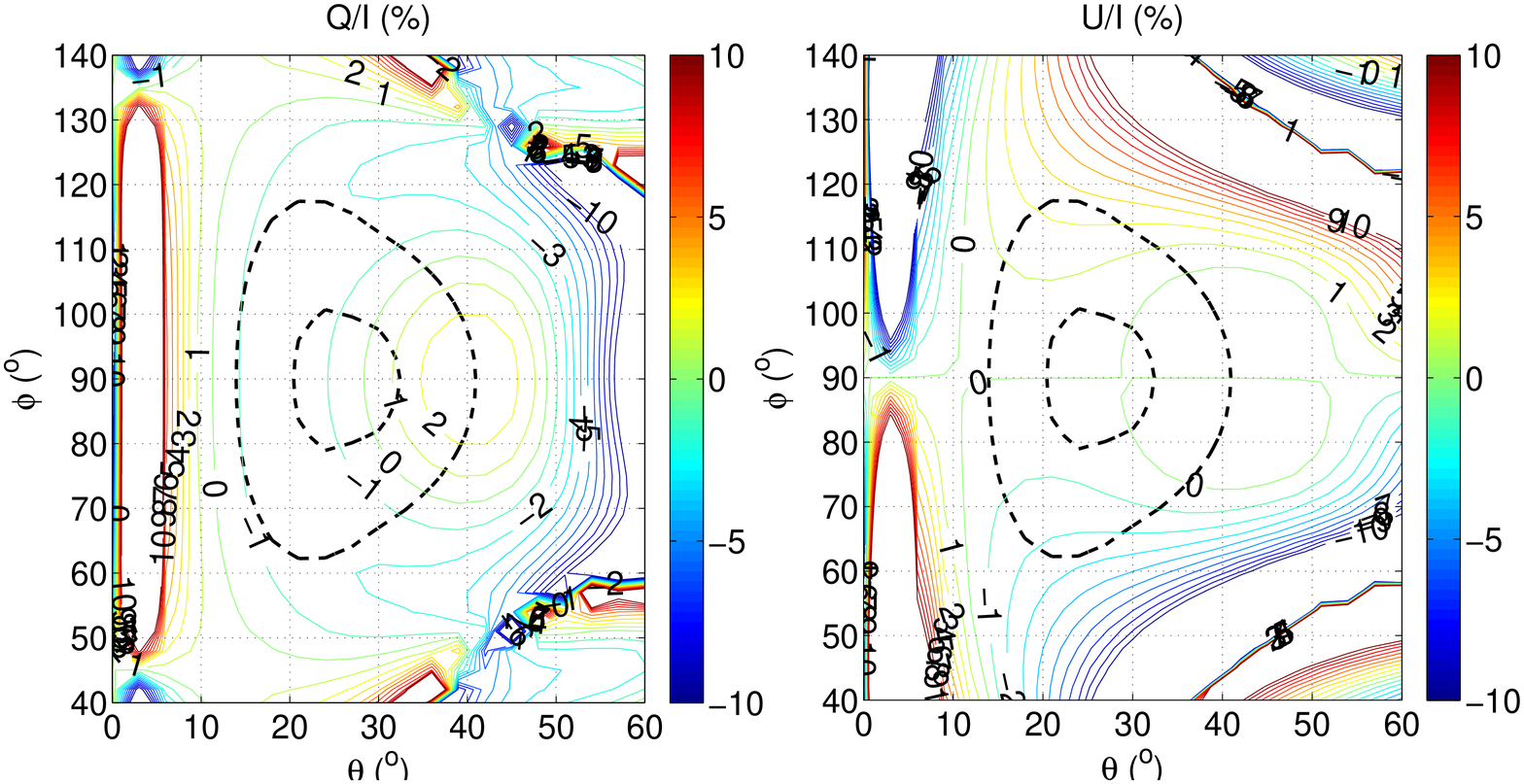}}
	\end{center}
\caption{Q and U Stokes leakages at 155~MHz for $Az=0^{o}, Za=28^{o}$  pointing after calibration with average embedded element pattern. Normalized power gain contours of 0.5 and 0.9 are superimposed on contour plots with 1\% step per contour.}
\label{fig:155_28_avg}
\end{figure*}

\begin{figure*}[htb]
	\begin{center}
	{\includegraphics[width=4.25in]{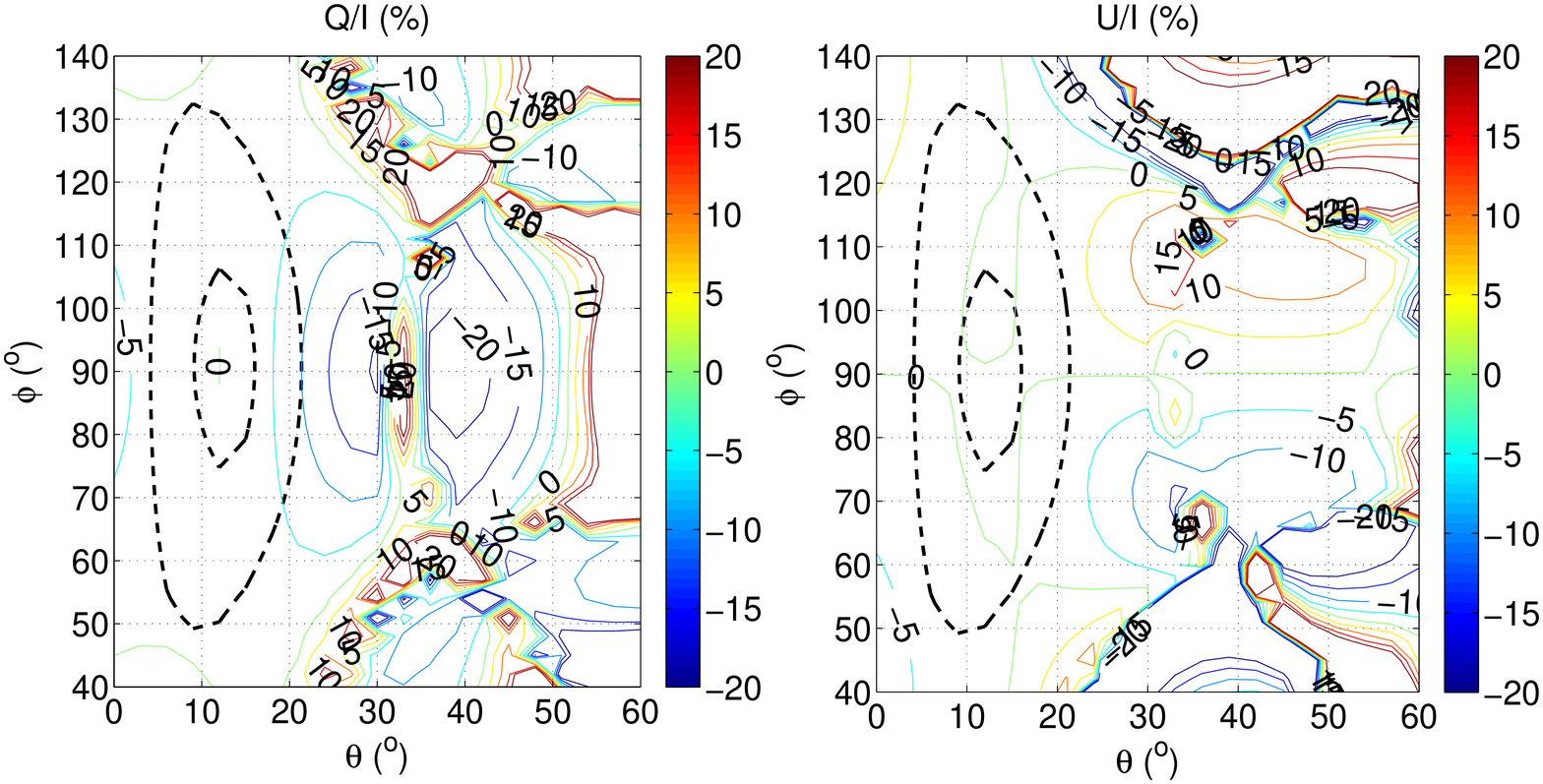}}
	\end{center}
\caption{Q and U Stokes leakages at 216~MHz for $Az=0^{o}, Za=14^{o}$  pointing after calibration with average embedded element pattern. Normalized power gain contours of 0.5 and 0.9 are superimposed on contour plots with 5\% step per contour.}
\label{fig:216_14_avg}
\end{figure*}

\begin{figure*}[htb]
	\begin{center}
	{\includegraphics[width=4.25in]{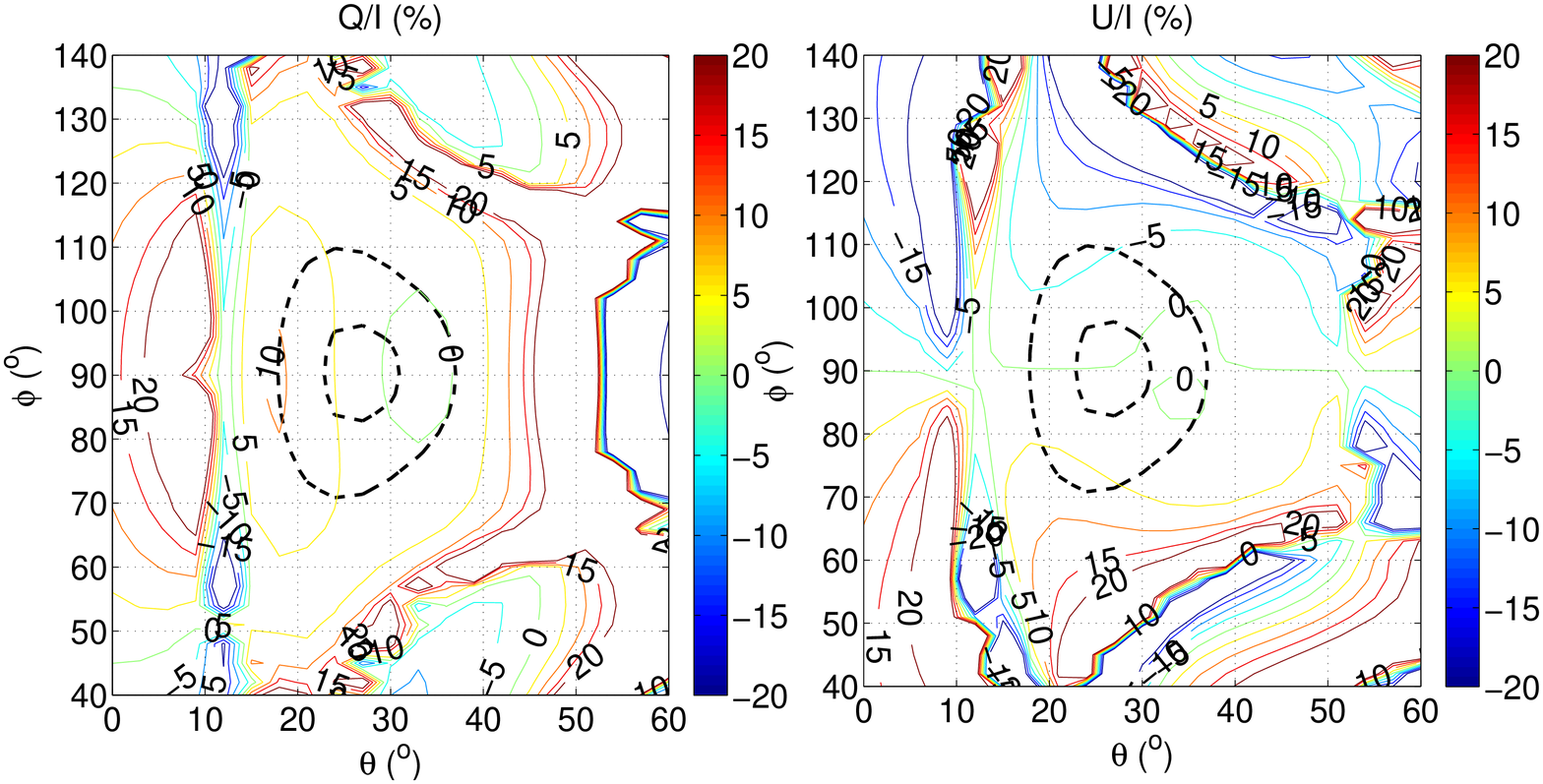}}
	\end{center}
\caption{Q and U Stokes leakages at 216~MHz for $Az=0^{o}, Za=28^{o}$  pointing after calibration with average embedded element pattern. Normalized power gain contours of 0.5 and 0.9 are superimposed on contour plots with 5\% step per contour.}
\label{fig:216_28_avg}
\end{figure*}

Again, we calculate residual Stokes leakage by taking (\ref{eqn:full}) as the ``real'' Jones matrix and (\ref{eqn:Javg}) as the Jones matrix estimate. The results are reported in Figs.~\ref{fig:155_28_avg}, \ref{fig:216_14_avg}, and \ref{fig:216_28_avg} and are summarized in Tab.~\ref{tab:hertz_avg}. To facilitate comparisons, we again point to plots of $Q/I$ on $Az=0^{\circ}$ plane at the same frequencies and pointing angles as shown in Figs.~\ref{fig:155_28_cut},~\ref{fig:216_14_cut} and~\ref{fig:216_28_cut} with legends ``avg.'' Improvement over the Hertzian dipole model is evident overall. The $Q/I$  leakages of the average embedded element pattern model tend to be centered closer to zero as zero contours cut through 0.9 normalized power beams in Figs.~\ref{fig:155_28_avg}, \ref{fig:216_14_avg}, and \ref{fig:216_28_avg} (similarly, improved centering is evident in Figs.~\ref{fig:155_28_cut},~\ref{fig:216_14_cut} and~\ref{fig:216_28_cut}).  Moreover, reduction in $U/I$ leakage in the main beam is obvious as shown in the bottom half of Tab.~\ref{tab:hertz_avg}. Thus, the average embedded element pattern should deliver noticeable improvement over current Hertzian dipole model, however, outliers in the order of up to 10\% in $|Q/I|$ may remain. Further improvement might be achieved through the full embedded element pattern model as discussed in the previous Subsection. 

\begin{table} [htb]
	\centering
	\begin{tabular}{ c c c c }
	 MHz & $Za~(^{\circ})$ & Hz $Q/I$ (\%)& avg. $Q/I (\%)$\\
	 \hline
	 155 & 28 & -1 to -7 & -1 to 3\\
	 216 & 14 & 0 to 16 & 0 to -10\\
	 216 & 28 & 30 & 0 to 10 \\
	&&&\\
	  MHz & $Za~(^{\circ})$ & Hz $U/I$ (\%)& avg. $U/I (\%)$\\
	 \hline
	 155 & 28 & 5 to -5 & 3 to -3\\
	 216 & 14 &  12 to -12 & 1 to -1\\
	 216 & 28 & 17 to -17 & 6 to -6\\
	\end{tabular}
	\caption{Comparison between calculated Stokes Q and U leakages using Hertzian dipole (Hz) and average embedded element pattern (avg.). The values are inferred from $Q/I$ and $U/I$ on the 0.5 normalized power gain contours in Figs.~\ref{fig:155_28_short}, \ref{fig:216_14_short}, \ref{fig:216_28_short} (Hz) and Figs.~\ref{fig:155_28_avg}, \ref{fig:216_14_avg}, \ref{fig:216_28_avg} (avg.).}
	\label{tab:hertz_avg}
\end{table}

\section{Conclusion}
\label{sec:concl}
We have learned interesting and useful lessons through this exercise. First, Jones matrix estimates that ignore mutual coupling will exhibit significant Stokes Q leakage (up to 20\%) at high frequencies (where element dimensions are comparable to or greater than wavelength) as imbalance in the ``X'' and ``Y'' port currents is unaccounted for. This is especially evident in the principal planes as the mutual coupling is asymmetric. Second, accurate high frequency polarimetry with LFAA requires precise modeling of both mutual coupling and (embedded) element effects. We are able to reproduce observation artifacts in $Q/I$ using the full embedded element pattern model with accuracy of approximately 2 to 5\%. This is indicative of the accuracy of the full embedded element model.  Finally, upgrading the Hertzian dipole model to one based on ``average'' embedded element pattern should deliver a noticeable improvement. Our calculations predict that $|Q/I|$ in the order of a few percent (with possible outliers up to approximately 10\%) is achievable.


%
%
%
%
%
%
%

\begin{acknowledgments}
We acknowledge the Wajarri Yamatji people as the traditional owners of the Murchison Radio Astronomy Observatory site. The MRO is operated
by CSIRO, whose assistance we acknowledge. The Centre for All-sky Astrophysics (CAASTRO) is an Australian Research Council Centre of Excellence, funded by grant CE110001020. Discussions on this topic with C. Jackson, J. Morgan, S. Ord, T. Colegate and D. Mitchell are gratefully acknowledged. Data on which figures and tables herein are based may be obtained by contacting the corresponding author Adrian Sutinjo (adrian.sutinjo@icrar.org) or Emil Lenc (elenc@me.com).
\end{acknowledgments}

\bibliographystyle{agufull08}

\begin{thebibliography}{}

\providecommand{\natexlab}[1]{#1}
\expandafter\ifx\csname urlstyle\endcsname\relax
  \providecommand{\doi}[1]{doi:\discretionary{}{}{}#1}\else
  \providecommand{\doi}{doi:\discretionary{}{}{}\begingroup
  \urlstyle{rm}\Url}\fi

\bibitem[{\textit{Balanis}(1989)}]{Balanis_1989_app}
Balanis, C.~A. (1989), \textit{Advanced Engineering Electromagnetics}, chap.
  Appendix, Wiley.

\bibitem[{\textit{Cheng}(1992)}]{Cheng_ch11}
Cheng, D.~K. (1992), \textit{Field and Waves Electromagnetics}, chap.~11, 2nd
  ed., Addison-Wesley.

\bibitem[{\textit{Coster et~al.}(2012)\textit{Coster, Herne, Erickson, and
  Oberoi}}]{RDS:RDS6017}
Coster, A., D.~Herne, P.~Erickson, and D.~Oberoi (2012), Using the murchison
  widefield array to observe midlatitude space weather, \textit{Radio Science},
  \textit{47}(6), n/a--n/a, \doi{10.1029/2012RS004993}.

\bibitem[{\textit{Ellingson et~al.}(2009)\textit{Ellingson, Clarke, Cohen,
  Craig, Kassim, Pihlstrom, Rickard, and Taylor}}]{5109716}
Ellingson, S., T.~Clarke, A.~Cohen, J.~Craig, N.~Kassim, Y.~Pihlstrom,
  L.~Rickard, and G.~Taylor (2009), The long wavelength array,
  \textit{Proceedings of the IEEE}, \textit{97}(8), 1421--1430,
  \doi{10.1109/JPROC.2009.2015683}.

\bibitem[{\textit{{Hamaker}}(2000{\natexlab{a}})}]{2000prat.conf..323H}
{Hamaker}, J. (2000{\natexlab{a}}), {Self-Calibration in the {SKA}: Dealing
  with Inherently Strong Instrumental Polarization}, in \textit{Perspectives on
  Radio Astronomy: Technologies for Large Antenna Arrays}, edited by A.~B.
  {Smolders} and M.~P. {van Haarlem}, p. 323.

\bibitem[{\textit{{Hamaker}}(2000{\natexlab{b}})}]{2000SPIE.4015..353H}
{Hamaker}, J.~P. (2000{\natexlab{b}}), {Self-calibration of arrays whose
  elements are strongly polarized}, in \textit{Radio Telescopes},
  \textit{Society of Photo-Optical Instrumentation Engineers (SPIE) Conference
  Series}, vol. 4015, edited by H.~R. {Butcher}, pp. 353--365.

\bibitem[{\textit{Hansen}(1998)}]{Hansen_1998_ch7}
Hansen, R.~C. (1998), \textit{Phased Array Antennas}, Wiley-Interscience.

\bibitem[{\textit{Helmboldt et~al.}(2014)\textit{Helmboldt, Ellingson, Hartman,
  Lazio, Taylor, Wilson, and Wolfe}}]{RDS:RDS20099}
Helmboldt, J.~F., S.~W. Ellingson, J.~M. Hartman, T.~J.~W. Lazio, G.~B. Taylor,
  T.~L. Wilson, and C.~N. Wolfe (2014), All-sky imaging of meteor trails at
  55.25 mhz with the first station of the long wavelength array, \textit{Radio
  Science}, \textit{49}(3), 157--180, \doi{10.1002/2013RS005220}.

\bibitem[{\textit{Hirasawa}(1992)}]{Hirasawa_1992_ch2}
Hirasawa, K. (1992), \textit{Analysis, Design, and Measurements of Small and
  Low-Profile Antennas}, chap.~2, Artech House.

\bibitem[{\textit{Kelley}(2002)}]{1016399}
Kelley, D. (2002), Relationships between active element patterns and mutual
  impedance matrices in phased array antennas, in \textit{Antennas and
  Propagation Society International Symposium, 2002. IEEE}, vol.~1, pp.
  524--527 vol.1, \doi{10.1109/APS.2002.1016399}.

\bibitem[{\textit{Kelley and Stutzman}(1993)}]{273305}
Kelley, D., and W.~Stutzman (1993), Array antenna pattern modeling methods that
  include mutual coupling effects, \textit{Antennas and Propagation, IEEE
  Transactions on}, \textit{41}(12), 1625--1632, \doi{10.1109/8.273305}.

\bibitem[{\textit{Kraus}(1988)}]{Kraus_1988_ch4}
Kraus, J. (1988), \textit{Antennas}, chap.~4, McGraw-Hill.

\bibitem[{\textit{Lonsdale et~al.}(2009)\textit{Lonsdale, Cappallo, Morales,
  Briggs, Benkevitch, Bowman, Bunton, Burns, Corey, deSouza, Doeleman, Derome,
  Deshpande, Gopala, Greenhill, Herne, Hewitt, Kamini, Kasper, Kincaid, Kocz,
  Kowald, Kratzenberg, Kumar, Lynch, Madhavi, Matejek, Mitchell, Morgan,
  Oberoi, Ord, Pathikulangara, Prabu, Rogers, Roshi, Salah, Sault, Shankar,
  Srivani, Stevens, Tingay, Vaccarella, Waterson, Wayth, Webster, Whitney,
  Williams, and Williams}}]{Lonsdale_2009}
Lonsdale, C., R.~Cappallo, M.~Morales, F.~Briggs, L.~Benkevitch, J.~Bowman,
  J.~Bunton, S.~Burns, B.~Corey, L.~deSouza, S.~Doeleman, M.~Derome,
  A.~Deshpande, M.~Gopala, L.~Greenhill, D.~Herne, J.~Hewitt, P.~Kamini,
  J.~Kasper, B.~Kincaid, J.~Kocz, E.~Kowald, E.~Kratzenberg, D.~Kumar,
  M.~Lynch, S.~Madhavi, M.~Matejek, D.~Mitchell, E.~Morgan, D.~Oberoi, S.~Ord,
  J.~Pathikulangara, T.~Prabu, A.~Rogers, A.~Roshi, J.~Salah, R.~Sault,
  N.~Shankar, K.~Srivani, J.~Stevens, S.~Tingay, A.~Vaccarella, M.~Waterson,
  R.~Wayth, R.~Webster, A.~Whitney, A.~Williams, and C.~Williams (2009), The
  murchison widefield array: Design overview, \textit{Proceedings of the IEEE},
  \textit{97}(8), 1497 --1506, \doi{10.1109/JPROC.2009.2017564}.

\bibitem[{\textit{McKinley}(2012)}]{McKinley_2012}
McKinley, B. (2012), Comparison of {MWA} tile beam patterns, \textit{{MWA}
  memo}, Australian National University.

\bibitem[{\textit{Ord et~al.}(2010)\textit{Ord, Mitchell, Wayth, Greenhill,
  Bernardi, Gleadow, Edgar, Clark, Allen, Arcus, Benkevitch, Bowman, Briggs,
  Bunton, Burns, Cappallo, Coles, Corey, deSouza, Doeleman, Derome, Deshpande,
  Emrich, Goeke, Gopalakrishna, Herne, Hewitt, Kamini, Kaplan, Kasper, Kincaid,
  Kocz, Kowald, Kratzenberg, Kumar, Lonsdale, Lynch, McWhirter, Madhavi,
  Matejek, Morales, Morgan, Oberoi, Pathikulangara, Prabu, Rogers, Roshi,
  Salah, Schinkel, Shankar, Srivani, Stevens, Tingay, Vaccarella, Waterson,
  Webster, Whitney, Williams, and Williams}}]{Ord2010}
Ord, S.~M., D.~A. Mitchell, R.~B. Wayth, L.~J. Greenhill, G.~Bernardi,
  S.~Gleadow, R.~G. Edgar, M.~A. Clark, G.~Allen, W.~Arcus, L.~Benkevitch,
  J.~D. Bowman, F.~H. Briggs, J.~D. Bunton, S.~Burns, R.~J. Cappallo, W.~A.
  Coles, B.~E. Corey, L.~deSouza, S.~S. Doeleman, M.~Derome, A.~Deshpande,
  D.~Emrich, R.~Goeke, M.~R. Gopalakrishna, D.~Herne, J.~N. Hewitt, P.~A.
  Kamini, D.~L. Kaplan, J.~C. Kasper, B.~B. Kincaid, J.~Kocz, E.~Kowald,
  E.~Kratzenberg, D.~Kumar, C.~J. Lonsdale, M.~J. Lynch, S.~R. McWhirter,
  S.~Madhavi, M.~Matejek, M.~F. Morales, E.~Morgan, D.~Oberoi,
  J.~Pathikulangara, T.~Prabu, A.~E.~E. Rogers, A.~Roshi, J.~E. Salah,
  A.~Schinkel, N.~U. Shankar, K.~S. Srivani, J.~Stevens, S.~J. Tingay,
  A.~Vaccarella, M.~Waterson, R.~L. Webster, A.~R. Whitney, A.~Williams, and
  C.~Williams (2010), Interferometric imaging with the 32 element murchison
  wide-field array, \textit{Publications of the Astronomical Society of the
  Pacific}, \textit{122}(897), pp. 1353--1366.

\bibitem[{\textit{Padhi}(2011)}]{SP_soil}
Padhi, S. (2011), Characteristic of conical wire spiral antenna with dispersive
  soil, \textit{Internal document}, ICRAR/Curtin University.

\bibitem[{\textit{Pozar}(1994)}]{310010}
Pozar, D. (1994), The active element pattern, \textit{Antennas and Propagation,
  IEEE Transactions on}, \textit{42}(8), 1176--1178, \doi{10.1109/8.310010}.

\bibitem[{\textit{Pozar}(1998)}]{Pozar_1998_ch4}
Pozar, D.~M. (1998), \textit{Microwave Engineering}, chap.~4, 2nd ed., Wiley.

\bibitem[{\textit{Smirnov}(2011)}]{Smirnov:2011vp}
Smirnov, O.~M. (2011), {Revisiting the radio interferometer measurement
  equation. I. A full-sky Jones formalism}, \textit{Astron.Astrophys.},
  \textit{527}, A106.

\bibitem[{\textit{Stutzman and Thiele}(1998)}]{Stutz_1998}
Stutzman, W.~L., and G.~A. Thiele (1998), \textit{Antenna Theory and Design},
  2nd ed., Wiley.

\bibitem[{\textit{Sutinjo and Hall}(2013)}]{6420896}
Sutinjo, A., and P.~Hall (2013), Intrinsic cross-polarization ratio of
  dual-linearly polarized antennas for low-frequency radio astronomy,
  \textit{Antennas and Propagation, IEEE Transactions on}, \textit{61}(5),
  2852--2856, \doi{10.1109/TAP.2013.2243101}.

\bibitem[{\textit{Sutinjo and Hall}(2014)}]{Sutinjo_TAP_TBP2014}
Sutinjo, A., and P.~Hall (2014), Antenna rotation error tolerance for a
  low-frequency aperture array polarimeter, \textit{Antennas and Propagation,
  IEEE Transactions on}, \textit{PP}(99), 1--1, \doi{10.1109/TAP.2014.2312201}.

\bibitem[{\textit{Taylor et~al.}(1999)\textit{Taylor, Carilli, and
  Perley}}]{taylor_synthesis_1999}
Taylor, G.~B., C.~L. Carilli, and R.~A. Perley (1999), Synthesis imaging in
  radio astronomy {II}.

\bibitem[{\textit{{Tingay} et~al.}(2013)\textit{{Tingay}, {Goeke}, {Bowman},
  {Emrich}, {Ord}, {Mitchell}, {Morales}, {Booler}, {Crosse}, {Wayth},
  {Lonsdale}, {Tremblay}, {Pallot}, {Colegate}, {Wicenec}, {Kudryavtseva},
  {Arcus}, {Barnes}, {Bernardi}, {Briggs}, {Burns}, {Bunton}, {Cappallo},
  {Corey}, {Deshpande}, {Desouza}, {Gaensler}, {Greenhill}, {Hall}, {Hazelton},
  {Herne}, {Hewitt}, {Johnston-Hollitt}, {Kaplan}, {Kasper}, {Kincaid},
  {Koenig}, {Kratzenberg}, {Lynch}, {Mckinley}, {Mcwhirter}, {Morgan},
  {Oberoi}, {Pathikulangara}, {Prabu}, {Remillard}, {Rogers}, {Roshi}, {Salah},
  {Sault}, {Udaya-Shankar}, {Schlagenhaufer}, {Srivani}, {Stevens},
  {Subrahmanyan}, {Waterson}, {Webster}, {Whitney}, {Williams}, {Williams}, and
  {Wyithe}}}]{2013PASA...30....7T}
{Tingay}, S.~J., R.~{Goeke}, J.~D. {Bowman}, D.~{Emrich}, S.~M. {Ord}, D.~A.
  {Mitchell}, M.~F. {Morales}, T.~{Booler}, B.~{Crosse}, R.~B. {Wayth}, C.~J.
  {Lonsdale}, S.~{Tremblay}, D.~{Pallot}, T.~{Colegate}, A.~{Wicenec},
  N.~{Kudryavtseva}, W.~{Arcus}, D.~{Barnes}, G.~{Bernardi}, F.~{Briggs},
  S.~{Burns}, J.~D. {Bunton}, R.~J. {Cappallo}, B.~E. {Corey}, A.~{Deshpande},
  L.~{Desouza}, B.~M. {Gaensler}, L.~J. {Greenhill}, P.~J. {Hall}, B.~J.
  {Hazelton}, D.~{Herne}, J.~N. {Hewitt}, M.~{Johnston-Hollitt}, D.~L.
  {Kaplan}, J.~C. {Kasper}, B.~B. {Kincaid}, R.~{Koenig}, E.~{Kratzenberg},
  M.~J. {Lynch}, B.~{Mckinley}, S.~R. {Mcwhirter}, E.~{Morgan}, D.~{Oberoi},
  J.~{Pathikulangara}, T.~{Prabu}, R.~A. {Remillard}, A.~E.~E. {Rogers},
  A.~{Roshi}, J.~E. {Salah}, R.~J. {Sault}, N.~{Udaya-Shankar},
  F.~{Schlagenhaufer}, K.~S. {Srivani}, J.~{Stevens}, R.~{Subrahmanyan},
  M.~{Waterson}, R.~L. {Webster}, A.~R. {Whitney}, A.~{Williams}, C.~L.
  {Williams}, and J.~S.~B. {Wyithe} (2013), {The Murchison Widefield Array: The
  Square Kilometre Array Precursor at Low Radio Frequencies},
  \textit{Publications of the Astronomical Society of Australia}, \textit{30},
  e007, \doi{10.1017/pasa.2012.007}.

\bibitem[{\textit{{van Haarlem, M. P.} et~al.}(2013)\textit{{van Haarlem, M.
  P.}, {Wise, M. W.}, {Gunst, A. W.}, {Heald, G.}, {McKean, J. P.}, {Hessels,
  J. W. T.}, {de Bruyn, A. G.}, {Nijboer, R.}, {Swinbank, J.}, {Fallows, R.},
  {Brentjens, M.}, {Nelles, A.}, {Beck, R.}, {Falcke, H.}, {Fender, R.},
  {Hörandel, J.}, {Koopmans, L. V. E.}, {Mann, G.}, {Miley, G.}, {Röttgering,
  H.}, {Stappers, B. W.}, {Wijers, R. A. M. J.}, {Zaroubi, S.}, {van den Akker,
  M.}, {Alexov, A.}, {Anderson, J.}, {Anderson, K.}, {van Ardenne, A.}, {Arts,
  M.}, {Asgekar, A.}, {Avruch, I. M.}, {Batejat, F.}, {Bähren, L.}, {Bell, M.
  E.}, {Bell, M. R.}, {van Bemmel, I.}, {Bennema, P.}, {Bentum, M. J.},
  {Bernardi, G.}, {Best, P.}, {Bîrzan, L.}, {Bonafede, A.}, {Boonstra, A.-J.},
  {Braun, R.}, {Bregman, J.}, {Breitling, F.}, {van de Brink, R. H.},
  {Broderick, J.}, {Broekema, P. C.}, {Brouw, W. N.}, {Brüggen, M.}, {Butcher,
  H. R.}, {van Cappellen, W.}, {Ciardi, B.}, {Coenen, T.}, {Conway, J.},
  {Coolen, A.}, {Corstanje, A.}, {Damstra, S.}, {Davies, O.}, {Deller, A. T.},
  {Dettmar, R.-J.}, {van Diepen, G.}, {Dijkstra, K.}, {Donker, P.}, {Doorduin,
  A.}, {Dromer, J.}, {Drost, M.}, {van Duin, A.}, {Eislöffel, J.}, {van Enst,
  J.}, {Ferrari, C.}, {Frieswijk, W.}, {Gankema, H.}, {Garrett, M. A.}, {de
  Gasperin, F.}, {Gerbers, M.}, {de Geus, E.}, {Grießmeier, J.-M.}, {Grit,
  T.}, {Gruppen, P.}, {Hamaker, J. P.}, {Hassall, T.}, {Hoeft, M.}, {Holties,
  H. A.}, {Horneffer, A.}, {van der Horst, A.}, {van Houwelingen, A.},
  {Huijgen, A.}, {Iacobelli, M.}, {Intema, H.}, {Jackson, N.}, {Jelic, V.}, {de
  Jong, A.}, {Juette, E.}, {Kant, D.}, {Karastergiou, A.}, {Koers, A.},
  {Kollen, H.}, {Kondratiev, V. I.}, {Kooistra, E.}, {Koopman, Y.}, {Koster,
  A.}, {Kuniyoshi, M.}, {Kramer, M.}, {Kuper, G.}, {Lambropoulos, P.}, {Law,
  C.}, {van Leeuwen, J.}, {Lemaitre, J.}, {Loose, M.}, {Maat, P.}, {Macario,
  G.}, {Markoff, S.}, {Masters, J.}, {McFadden, R. A.}, {McKay-Bukowski, D.},
  {Meijering, H.}, {Meulman, H.}, {Mevius, M.}, {Middelberg, E.}, {Millenaar,
  R.}, {Miller-Jones, J. C. A.}, {Mohan, R. N.}, {Mol, J. D.}, {Morawietz, J.},
  {Morganti, R.}, {Mulcahy, D. D.}, {Mulder, E.}, {Munk, H.}, {Nieuwenhuis,
  L.}, {van Nieuwpoort, R.}, {Noordam, J. E.}, {Norden, M.}, {Noutsos, A.},
  {Offringa, A. R.}, {Olofsson, H.}, {Omar, A.}, {Orrú, E.}, {Overeem, R.},
  {Paas, H.}, {Pandey-Pommier, M.}, {Pandey, V. N.}, {Pizzo, R.}, {Polatidis,
  A.}, {Rafferty, D.}, {Rawlings, S.}, {Reich, W.}, {de Reijer, J.-P.},
  {Reitsma, J.}, {Renting, G. A.}, {Riemers, P.}, {Rol, E.}, {Romein, J. W.},
  {Roosjen, J.}, {Ruiter, M.}, {Scaife, A.}, {van der Schaaf, K.}, {Scheers,
  B.}, {Schellart, P.}, {Schoenmakers, A.}, {Schoonderbeek, G.}, {Serylak, M.},
  {Shulevski, A.}, {Sluman, J.}, {Smirnov, O.}, {Sobey, C.}, {Spreeuw, H.},
  {Steinmetz, M.}, {Sterks, C. G. M.}, {Stiepel, H.-J.}, {Stuurwold, K.},
  {Tagger, M.}, {Tang, Y.}, {Tasse, C.}, {Thomas, I.}, {Thoudam, S.}, {Toribio,
  M. C.}, {van der Tol, B.}, {Usov, O.}, {van Veelen, M.}, {van der Veen,
  A.-J.}, {ter Veen, S.}, {Verbiest, J. P. W.}, {Vermeulen, R.}, {Vermaas, N.},
  {Vocks, C.}, {Vogt, C.}, {de Vos, M.}, {van der Wal, E.}, {van Weeren, R.},
  {Weggemans, H.}, {Weltevrede, P.}, {White, S.}, {Wijnholds, S. J.},
  {Wilhelmsson, T.}, {Wucknitz, O.}, {Yatawatta, S.}, {Zarka, P.}, {Zensus,
  A.}, and {van Zwieten, J.}}}]{refId0}
{van Haarlem, M. P.}, {Wise, M. W.}, {Gunst, A. W.}, {Heald, G.}, {McKean, J.
  P.}, {Hessels, J. W. T.}, {de Bruyn, A. G.}, {Nijboer, R.}, {Swinbank, J.},
  {Fallows, R.}, {Brentjens, M.}, {Nelles, A.}, {Beck, R.}, {Falcke, H.},
  {Fender, R.}, {Hörandel, J.}, {Koopmans, L. V. E.}, {Mann, G.}, {Miley, G.},
  {Röttgering, H.}, {Stappers, B. W.}, {Wijers, R. A. M. J.}, {Zaroubi, S.},
  {van den Akker, M.}, {Alexov, A.}, {Anderson, J.}, {Anderson, K.}, {van
  Ardenne, A.}, {Arts, M.}, {Asgekar, A.}, {Avruch, I. M.}, {Batejat, F.},
  {Bähren, L.}, {Bell, M. E.}, {Bell, M. R.}, {van Bemmel, I.}, {Bennema, P.},
  {Bentum, M. J.}, {Bernardi, G.}, {Best, P.}, {Bîrzan, L.}, {Bonafede, A.},
  {Boonstra, A.-J.}, {Braun, R.}, {Bregman, J.}, {Breitling, F.}, {van de
  Brink, R. H.}, {Broderick, J.}, {Broekema, P. C.}, {Brouw, W. N.}, {Brüggen,
  M.}, {Butcher, H. R.}, {van Cappellen, W.}, {Ciardi, B.}, {Coenen, T.},
  {Conway, J.}, {Coolen, A.}, {Corstanje, A.}, {Damstra, S.}, {Davies, O.},
  {Deller, A. T.}, {Dettmar, R.-J.}, {van Diepen, G.}, {Dijkstra, K.}, {Donker,
  P.}, {Doorduin, A.}, {Dromer, J.}, {Drost, M.}, {van Duin, A.}, {Eislöffel,
  J.}, {van Enst, J.}, {Ferrari, C.}, {Frieswijk, W.}, {Gankema, H.}, {Garrett,
  M. A.}, {de Gasperin, F.}, {Gerbers, M.}, {de Geus, E.}, {Grießmeier,
  J.-M.}, {Grit, T.}, {Gruppen, P.}, {Hamaker, J. P.}, {Hassall, T.}, {Hoeft,
  M.}, {Holties, H. A.}, {Horneffer, A.}, {van der Horst, A.}, {van
  Houwelingen, A.}, {Huijgen, A.}, {Iacobelli, M.}, {Intema, H.}, {Jackson,
  N.}, {Jelic, V.}, {de Jong, A.}, {Juette, E.}, {Kant, D.}, {Karastergiou,
  A.}, {Koers, A.}, {Kollen, H.}, {Kondratiev, V. I.}, {Kooistra, E.},
  {Koopman, Y.}, {Koster, A.}, {Kuniyoshi, M.}, {Kramer, M.}, {Kuper, G.},
  {Lambropoulos, P.}, {Law, C.}, {van Leeuwen, J.}, {Lemaitre, J.}, {Loose,
  M.}, {Maat, P.}, {Macario, G.}, {Markoff, S.}, {Masters, J.}, {McFadden, R.
  A.}, {McKay-Bukowski, D.}, {Meijering, H.}, {Meulman, H.}, {Mevius, M.},
  {Middelberg, E.}, {Millenaar, R.}, {Miller-Jones, J. C. A.}, {Mohan, R. N.},
  {Mol, J. D.}, {Morawietz, J.}, {Morganti, R.}, {Mulcahy, D. D.}, {Mulder,
  E.}, {Munk, H.}, {Nieuwenhuis, L.}, {van Nieuwpoort, R.}, {Noordam, J. E.},
  {Norden, M.}, {Noutsos, A.}, {Offringa, A. R.}, {Olofsson, H.}, {Omar, A.},
  {Orrú, E.}, {Overeem, R.}, {Paas, H.}, {Pandey-Pommier, M.}, {Pandey, V.
  N.}, {Pizzo, R.}, {Polatidis, A.}, {Rafferty, D.}, {Rawlings, S.}, {Reich,
  W.}, {de Reijer, J.-P.}, {Reitsma, J.}, {Renting, G. A.}, {Riemers, P.},
  {Rol, E.}, {Romein, J. W.}, {Roosjen, J.}, {Ruiter, M.}, {Scaife, A.}, {van
  der Schaaf, K.}, {Scheers, B.}, {Schellart, P.}, {Schoenmakers, A.},
  {Schoonderbeek, G.}, {Serylak, M.}, {Shulevski, A.}, {Sluman, J.}, {Smirnov,
  O.}, {Sobey, C.}, {Spreeuw, H.}, {Steinmetz, M.}, {Sterks, C. G. M.},
  {Stiepel, H.-J.}, {Stuurwold, K.}, {Tagger, M.}, {Tang, Y.}, {Tasse, C.},
  {Thomas, I.}, {Thoudam, S.}, {Toribio, M. C.}, {van der Tol, B.}, {Usov, O.},
  {van Veelen, M.}, {van der Veen, A.-J.}, {ter Veen, S.}, {Verbiest, J. P.
  W.}, {Vermeulen, R.}, {Vermaas, N.}, {Vocks, C.}, {Vogt, C.}, {de Vos, M.},
  {van der Wal, E.}, {van Weeren, R.}, {Weggemans, H.}, {Weltevrede, P.},
  {White, S.}, {Wijnholds, S. J.}, {Wilhelmsson, T.}, {Wucknitz, O.},
  {Yatawatta, S.}, {Zarka, P.}, {Zensus, A.}, and {van Zwieten, J.} (2013),
  {LOFAR}: The {LOw-Frequency ARray}, \textit{A\&A}, \textit{556}, A2,
  \doi{10.1051/0004-6361/201220873}.

\bibitem[{\textit{Wijnholds et~al.}(2011)\textit{Wijnholds, Bregman, and van
  Ardenne}}]{RDS:RDS5907}
Wijnholds, S.~J., J.~D. Bregman, and A.~van Ardenne (2011), Calibratability and
  its impact on configuration design for the {LOFAR} and {SKA} phased array
  radio telescopes, \textit{Radio Science}, \textit{46}(5), n/a--n/a,
  \doi{10.1029/2011RS004733}.


%
%
%

\end{thebibliography}

\end{article}

\end{document}